\documentclass[a4paper,fleqn,usenatbib]{mnras}
\usepackage{newtxtext,newtxmath}
\usepackage[T1]{fontenc}
\usepackage{ae,aecompl}

\usepackage{graphicx}	
\usepackage{amsmath}	
\usepackage{amssymb}	
\hypersetup{colorlinks=true,linkcolor=blue,citecolor=blue,filecolor=blue,urlcolor=blue,draft=false}
\usepackage{xspace}
\usepackage[per-mode=reciprocal,range-phrase=--,free-standing-units=true,use-xspace=true,range-units=single]{siunitx}
\usepackage[usenames,dvipsnames,svgnames,table]{xcolor}

\newcommand{\hi}{\ifmmode {\mathrm{H\,\textsc{i}}} \else H\,\textsc{i} \fi}
\newcommand{\hii}{\ifmmode {\mathrm{H\,\textsc{ii}}} \else H\,\textsc{ii} \fi}
\newcommand{\hei}{\ifmmode {\mathrm{He\,\textsc{i}}} \else He\,\textsc{i} \fi}
\newcommand{\heii}{\ifmmode {\mathrm{He\,\textsc{ii}}} \else He\,\textsc{ii} \fi}
\newcommand{\heiii}{\ifmmode {\mathrm{He\,\textsc{iii}}} \else He\,\textsc{iii} \fi}
\newcommand{\Mvir}{\ifmmode {M_{\rm vir}} \else $M_{\rm vir}$\xspace\fi}
\newcommand{\Mstar}{\ifmmode {M_\star} \else $M_{\star}$\xspace\fi}
\newcommand{\Mbh}{\ifmmode {M_\bullet} \else $M_{\bullet}$\xspace\fi}
\newcommand{\Lbol}{\ifmmode {L_{\rm bol}} \else $L_{\rm bol}$\xspace\fi}
\newcommand{\fedd}{\ifmmode {f_{\rm Edd}} \else $f_{\rm Edd}$\xspace\fi}
\newcommand{\NH}{\ifmmode {N_{\rm H}} \else $N_{\rm H}$\xspace\fi}

\newcommand{\Rvir}{\ifmmode {R_{\rm vir}} \else $R_{\rm vir}$\xspace\fi}
\newcommand{\fesc}{\ifmmode {f_{\rm esc}^{\star}} \else $f_{\rm esc}^{\star}$\xspace\fi}
\newcommand{\fescAGN}{\ifmmode {f_{\rm esc}^{\rm AGN}} \else $f_{\rm esc}^{\rm AGN}$\xspace\fi}
\newcommand{\Mgal}{\ifmmode {{\rm M}_{1500}} \else ${\rm M}_{1500}$\xspace\fi}
\newcommand{\Magn}{\ifmmode {M_{1450}} \else $M_{1450}$\xspace\fi}
\newcommand{\MUV}{\ifmmode {{\rm M}_{\rm UV}} \else ${\rm M}_{\rm UV}$\xspace\fi}
\newcommand{\lya}{\ifmmode {{\rm Ly}\alpha} \else ${\rm Ly}\alpha$\xspace\fi}

\DeclareSIUnit\year{yr}
\DeclareSIUnit\Msun{\ifmmode {\rm M}_{\odot} \else ${\rm M}_\odot$\xspace\fi}

\newcommand{\ramses}{\textsc{Ramses}\xspace}
\newcommand{\ramsesrt}{\textsc{Ramses-RT}\xspace}

\title[Bright galaxy and faint AGN at $z\sim 6$]{Modelling a bright $z=6$ galaxy at the faint end of the AGN luminosity function}

\author[M. Trebitsch et al.]{
Maxime Trebitsch,$^{1,2,3}$\thanks{E-mail: maxime.trebitsch@iap.fr}
Marta Volonteri$^{1}$
and Yohan Dubois$^{1}$
\\
$^{1}$Sorbonne Universit\'{e}, CNRS, UMR 7095, Institut d'Astrophysique de Paris, 98 bis bd Arago, 75014 Paris, France \\
$^{2}$Max-Planck-Institut f{\"u}r Astronomie, K{\"o}nigstuhl 17, 69117 Heidelberg, Germany \\
$^{3}$Zentrum f{\"u}r Astronomie der Universität Heidelberg, Institut f{\"u}r Theoretische Astrophysik, Albert-Ueberle-Str. 2, 69120 Heidelberg, Germany
}

\date{Accepted 2020 April 06. Received 2020 April 03; in original form 2019 July 16}

\pubyear{2019}

\begin{document}
\label{firstpage}
\pagerange{\pageref{firstpage}--\pageref{lastpage}}
\maketitle

\begin{abstract}
  Recent deep surveys have unravelled a population of faint active galactic nuclei (AGN) in the high redshift Universe, leading to various discussions on their nature and their role during the Epoch of Reionization. We use cosmological radiation-hydrodynamics simulations of a bright galaxy at $z \sim 6$ ($\Mstar \gtrsim \SI{e10}{\Msun}$) hosting an actively growing super-massive black hole to study the properties of these objects. In particular, we study how the black hole and the galaxy co-evolve and what is the relative contribution of the AGN and of the stellar populations to the luminosity budget of the system.
  We find that the feedback from the AGN has no strong effect on the properties of the galaxy, and does not increase the total ionizing luminosity of the host. The average escape fraction of our galaxy is around $f_{\rm esc} \sim 5\%$.
  While our galaxy would be selected as an AGN in deep X-ray surveys, most of the UV luminosity is originating from stellar populations.
  This confirms that there is a transition in the galaxy population from star forming galaxies to quasar hosts, with bright Lyman-Break Galaxies (LBGs) with \MUV around -22 falling in the overlap region. Our results also suggest that faint AGN do not contribute significantly to reionizing the Universe.
\end{abstract}




\section{Introduction}
\label{sec:intro}

Cosmic reionization is the process through which the initially neutral intergalactic medium (IGM) becomes ionized by sources of hard ultraviolet (UV) radiation during the first billion years of the Universe ($z \sim 20-6$).
The bulk of these photons is thought to be predominantly produced by massive stars in faint star forming galaxies \citep[e.g.][]{Robertson2015, Finkelstein2019}. In this picture, quasars mainly maintain the post-reionization UV background \citep[e.g.][]{Becker2013,Kulkarni2019}, with some additional role in determining the patchiness of the end of the process of reionization \citep{Chardin2015, Chardin2017,Kakiichi2018}. This is for instance supported by observations of faint lensed galaxies behind clusters, indicating that the faint-end of the galaxy UV luminosity function (LF) might be steep, with no sign of any turn-over brighter than $\MUV \lesssim -15$ \citep[e.g.][but see also \citealt{Atek2018} for a detailed analysis of the model uncertainties]{Bouwens2017, Livermore2017, Ishigaki2018}, and therefore that the number of faint galaxies able to produce ionizing radiation is large enough to reionize the Universe. In the meantime, modern radiation hydrodynamics (RHD) cosmological simulations suggest that actively star-forming galaxies hosted in low-mass dark matter (DM) haloes can provide enough ionizing radiation to reionize the Universe by $z\sim 6$ \citep[e.g.][]{Gnedin2014, Ocvirk2016, Rosdahl2018}.

In the recent years, a lot of work has been dedicated to identify faint active galactic nuclei (AGN) in the high-redshift Universe and determine their contribution to the reionization of the Universe \citep[e.g.][]{Giallongo2015, Ricci2017, Boutsia2018, Parsa2018, Matsuoka2018, Stevans2018}. These AGN, while less luminous than bright quasars, could in principle be far more numerous. For this reason, they have been suggested as an additional source of ionizing photons that could potentially play a significant role in reionizing the Universe. Understanding the properties of these objects is therefore highly relevant to the study of the sources of reionization: if, like bright quasars, faint AGN have a very high escape fraction as postulated e.g. by \citet{Giallongo2015}, they could significantly contribute to the reionization of the Universe. Conversely, if they are heavily obscured, or if a non-negligible fraction of their UV and ionizing luminosities is produced by the stellar populations of their host, their relevance to the reionization history would be greatly diminished.

In this work, we aim at studying objects in the intermediate regime between galaxies (fainter than $\MUV \gtrsim -22$) and quasars (brighter than $\MUV \lesssim -26$).
The properties of faint AGN and their host are still unclear. For instance, while the observations of \citet{Cristiani2016} suggest that the AGN escape fraction \fescAGN can reach high values for bright quasars, it is virtually unknown at the faint end. \citet{Grazian2018} find a high relative \fescAGN for their sample at $z \sim 4$, while the analysis of a sample of faint AGN selected in the SSA22 protocluster by \citet{Micheva2017} is suggestive of \fescAGN being below unity at $z\sim 3$ (although the sample size is small). \citet{Guaita2016} report the observation of one object with high \fescAGN, but several other objects in their sample have only only upper limits. Overall, while this could be suggestive that \fescAGN in the low-luminosity regime is well below unity, this is clearly not a solved problem. Independently of the value of \fescAGN, nuclear activity has been proposed as a solution to enhance the (stellar) escape fraction from the galaxy \fesc \citep{Seiler2018}. This scenario would be very challenging to test directly through observations, but can be investigated through dedicated RHD simulations.

The number density of faint AGN is close to that of the brightest Lyman-break galaxies (LBGs) observed, with the faint end of the AGN LF overlapping with the bright end of the galaxy LF as found by the SHELLQs \citep[e.g.][]{Matsuoka2018}, GOLDRUSH \citep{Ono2018} or SHELA survey \citep{Stevans2018}.
This overlap happens just around the luminosity regime probed by the objects investigated by \citet{Giallongo2015} at $z \sim 4$, who found in the COSMOS field a larger than expected number of AGN candidates of this magnitude or fainter. This echoes the results of \citet{Volonteri2017}, who found that in this regime, both the stellar populations and the nuclear activity contribute to the (UV) luminosity. The brightest of the $z\sim 6$ \lya emitters (LAEs) known to this date, such as \emph{Himiko} \citep{Ouchi2009} or CR7 and VR7 \citep{Matthee2017}, all have UV luminosities comparable to these faint AGN. Their observed properties are sometimes hard to explain with standard stellar populations or even population III stars \citep[see e.g.][for CR7]{Sobral2015,Hartwig2016}, but can be more consistent with an hidden AGN \citep{Bowler2017}. This would be consistent with the results of \citet{Hatfield2018}, who suggest that the bright LBGs observed at high-$z$ are not just particularly star forming but otherwise low mass galaxies, but are intrinsically massive objects hosted in $M_{\rm halo} \sim \SI{e11.5}{\Msun}$ haloes: galaxies in such haloes are expected to host a massive black hole (MBH) growing at their centre.

In this paper, we perform a series of high resolution radiation-hydrodynamical cosmological zoom simulations of a massive galaxy around the knee of the galaxy mass function at $z \sim 6$, with the goal of studying the properties of massive black holes living in actively growing galaxies. In particular, we aim at connecting the growth of the MBH to that of the galaxy, and conversely assessing how the active galactic nuclei (AGN) powered by the accretion onto the MBH affects the star formation history of the galaxy. In the context of the contribution of AGN to cosmic reionization, we want to quantify how the radiation produced by the AGN escapes in the IGM, and how the nuclear activity affects the escape of (stellar) ionizing radiation.
We first describe the simulations used in this work in Sect.~\ref{sec:method}. We then go on to present the properties of the galaxy and its central BH (Sect.~\ref{sec:BH-galaxy-coevolution}), how much ionizing radiation is produced by the system (Sect.~\ref{sec:effect-agn-escape}), and whether the object should be classified as a galaxy or an AGN (Sect.~\ref{sec:galaxy-agn-lumin}).

\section{Galaxy simulations}
\label{sec:method}

We use a set of zoom-in simulations performed with \ramsesrt, the RHD version of the public adaptive mesh refinement (AMR) code \ramses \footnote{\url{https://bitbucket.org/rteyssie/ramses/}} \citep{Teyssier2002, Rosdahl2013, Rosdahl2015}, already partially described in \citet{Trebitsch2019}. In this section, we briefly summarize the main features of the code and sub-grid models employed, and we refer the interested reader to \citet{Trebitsch2019} for a more detailed description.
\begin{figure*} 
  \includegraphics[width=.9\linewidth]{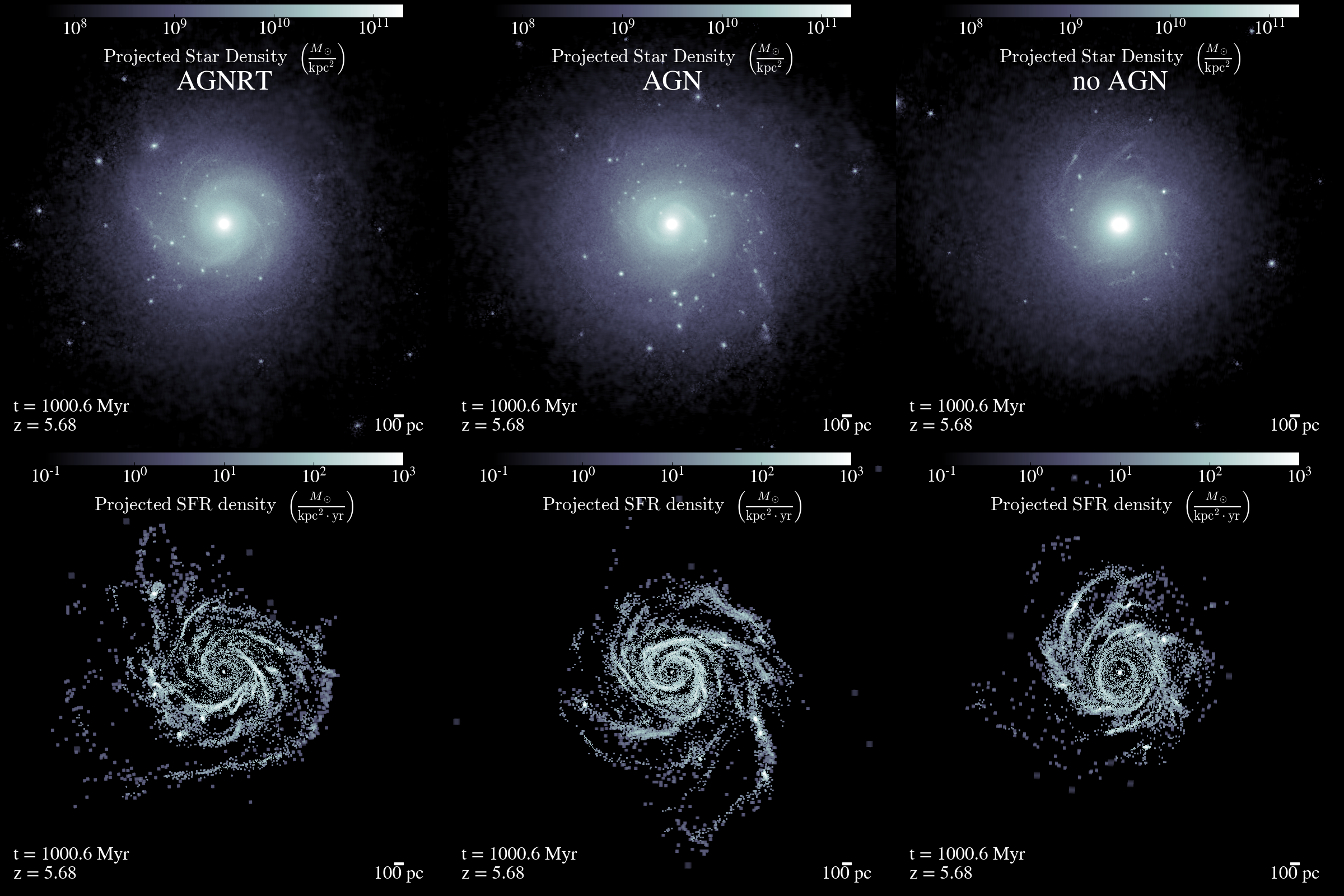}
  \caption{Illustration of the main galaxy at $z\sim 5.7$ for the three runs discussed in this work. The upper row shows the stellar mass surface density while the lower row shows the SFR surface density. The ``AGNRT'' run is on the left, the ``AGN'' run in the middle, and the run with ``no AGN'' on the right.}
  \label{fig:all_galaxies}
\end{figure*}

The collisionless particles (stars and dark matter) are evolved using a particle-mesh method with a cloud-in-cell interpolation. For the gas, \ramses solves the Euler equations with the second-order MUSCL scheme \citep{vanLeer1979} using the HLLC Riemann solver from \citet{Toro1994} and a MinMod total variation diminishing scheme to reconstruct the intercell-fluxes. For all simulations, we impose a Courant factor of 0.8 to define the timestep.

The AMR grid is refined using a quasi-Lagrangian criterion: a cell is selected for refinement if $\rho_{\rm DM} \Delta x^3 + (\Omega_{\rm DM}/\Omega_b)\rho_{\rm gas} \Delta x^3+ (\Omega_{\rm DM}/\Omega_b) \rho_* \Delta x^3 > 8\ m_{\rm DM}^{\rm HR}$, where $\rho_{\rm DM}$, $\rho_{\rm gas}$ and $\rho_*$ are respectively the DM, gas and stellar densities in the cell, $\Omega_{\rm DM}$ and $\Omega_b$ respectively the cosmic DM and baryon mass density, $\Delta x$ is the cell size, and $m_{\rm DM}^{\rm HR}$ is the mass of the highest resolution DM particle. In a DM-only run, this criterion would allow refinement as soon as there are at least 8 high-resolution DM particles in a cell.

\subsection{Initial conditions}
\label{sec:sims:ics}
We zoom on the galaxy described in \citet{Trebitsch2019}, which lives in a halo reaching a mass of $\sim \SI{3e11}{\Msun}$ at redshift $z \sim 5.7$ embedded in a cosmological volume of $40 h^{-1}$ comoving Mpc on a side. The initial conditions for both the initial DM-only run and the zoom region have been generated with \textsc{Music}\footnote{\url{https://bitbucket.org/ohahn/music/}} \citep{Hahn2011}, assuming a flat $\Lambda$CDM cosmology consistent with the \emph{Planck} results \citep[dark energy density $\Omega_\Lambda = 0.692$, total matter density $\Omega_m = 0.308$, Hubble parameter $h = 0.6781$ and baryon matter density $\Omega_b = 0.048$,][]{Planck2015}. We select the target halo in the final output with \textsc{HaloMaker} \citep{Tweed2009}, which uses the \textsc{AdaptaHOP} algorithm \citep{Aubert2004}.

The zoom region has an effective resolution of $4096^3$ elements (level $\ell = 12$), which translates in a mass resolution of $m_{\rm DM}^{\rm HR} \simeq \SI{e5}{\Msun}$ for the high-resolution particles.
For the RHD run, we then allow for refinement down to a minimum cell size of $\Delta x = 40 h^{-1}\mbox{Mpc}/2^{23} \simeq 7\,\mbox{pc}$. The gas in the initial conditions is assumed to be neutral and homogeneously metal poor, with an initial gas phase metallicity $Z = 5\times 10^{-3} Z_\odot = 10^{-4}$.

\subsection{Radiative transfer}
\label{sec:sims:radiative-transfer}
The RT module propagates the radiation emitted by both stars and BHs in three frequency intervals, accounting for the \hi-, \hei- and \heii-ionizing radiation fields. The radiation is then evolved on the AMR grid using a first-order Godunov method to solve the first two moments of the RT equation and assuming the M1 closure \citep{Levermore1984, Dubroca1999} for the Eddington tensor. We use the reduced speed of light approximation \citep{Gnedin2001, Rosdahl2013} to limit the cost of the simulation, with a reduced speed of light of $\tilde{c} = 0.01 c$.
The radiation is coupled to the gas through non-equilibrium thermochemistry for hydrogen and helium, assuming the on-the-spot approximation, and we ignore the radiation pressure exerted by the ionizing radiation on the gas. We discuss this assumption in Sect.~\ref{sec:ccl}.
Radiation is emitted by each star particle as a function of its age and metallicity following the models of \citet{Bruzual2003}, and from each AGN as a function of the BH mass and accretion rate. For the AGN radiation, we only release photons when the AGN is in ``quasar mode'' (see Sect.~\ref{sec:sims:bhagn}), and the spectrum follows a piecewise power-law  corresponding to a \citet{Shakura1973} thin disc extended by a power-law at high energy, with slope $\alpha = -1.7$ \citep{Lusso2015}. We normalize the spectrum by assuming that only a fraction $(1-f_{\rm IR})$ of the bolometric luminosity \Lbol escapes the inner dusty region, so that $f_{\rm IR} = 30\%$ of \Lbol is absorbed by dust and re-emitted as IR radiation, that we do not model here.

\subsection{Star formation and feedback}
\label{sec:sims:sffb}
At the resolution of our simulation, we describe the stars as particles with mass $m_\star \sim \SI{1.8e4}{\Msun}$ representing a single stellar population.
Star formation is modelled with a Schmidt-like law \citep{Schmidt1959}, with an approach similar to that \citet{Rasera2006}, but assuming a local star formation efficiency $\epsilon(\rho, c_s, \mathcal{M})$ computed the following `multi-ff PN' model of \citet{Federrath2012, Padoan2011}, where $\rho$ is the local density, $c_s$ the local sound speed, and $\mathcal{M}$ the local turbulent Mach number.
We only consider cells to be star forming when the local density $\rho$ exceeds a threshold\footnote{Our choice of $m_\star$ forces cells at the highest level to only form stars if $\rho \gtrsim \SI{1000}{\per\cubic\cm}$. The threshold prevents stars to form out of the high-resolution region, as the star formation is regulated by the local efficiency.} $\rho_0 = 1\,\mbox{cm}^{-3}$ (chosen as the typical ISM density), and when the local turbulent Mach number exceeds $\mathcal{M} \geq 2$.

We include feedback from massive stars through radiative feedback resulting from photoionization heating and type II supernovae (SNe). Photoionization heating is directly modelled in the simulation through the coupling between the ionizing photon field and the gas (see Sect.~\ref{sec:sims:radiative-transfer}).
For the SN feedback, we use the model of \citet{Kimm2014,Kimm2015}, which deposits mass and momentum in every cell around a star particle in a single event $t_{\rm SN} = 5\,\mbox{Myr}$ after it is formed. The amount of momentum released depends on the local density and metallicity of each neighbouring cell in order to capture correctly the momentum transfer at all stages of the Sedov blast wave.
Following \citet{Kimm2017}, we increase the final radial momentum from SNe when the Str\"omgren sphere of a star particle is unresolved, as suggested by \citet{Geen2015}.

\subsection{BH model}
\label{sec:sims:bhagn}

The BH seeding, growth and  associated feedback follow the fiducial implementation of \citet{Dubois2012}.
We represent SMBHs using sink particles with initial mass $M_{\bullet,0} = \SI{3e4}{\Msun}$. These sink particles are created in cells where both the gas and stellar density exceeds a threshold that we choose to be $\rho_{\rm sink} = 100\,\mbox{cm}^{-3}$, where the gas is Jeans-unstable, and where there is enough gas in the cell to form the sink particle.
Additionally, if there is a SMBH within $r_{\rm excl} = 40\,\mbox{kpc}$ of a selected cell, we block BH formation to exclude the formation of multiple SMBH in the same galaxy. Each sink particle is then surrounded by tracers in the form of massless `cloud' particles equally spaced by $\Delta x/2$ within a sphere of radius $4 \Delta x$ and moving with the SMBH, providing a convenient way to probe the gas properties around the BH.

BHs accrete gas following the classical Bondi-Hoyle-Lyttleton prescription \citep{Bondi1952}, $\dot{M}_{\rm BHL} = 4\pi G^2 \Mbh^2 \bar{\rho}/(\bar{c}_s^2 + \bar{v}_{\rm rel}^2)^{3/2}$, where $\Mbh$ is the BH mass, $\bar{\rho}$, $\bar{c}_s$, and $\bar{v}_{\rm rel}$ are respectively the average gas density, sound speed, and relative velocity between the BH and the surrounding gas. The bar notation denotes an averaging over the cloud particles.
We do not use any artificial boost for the gas accretion onto the BH. The accretion rate is limited to the value that produces the Eddington luminosity assuming a radiative efficiency of  $\epsilon_r = 0.1$, with   $L_{\rm Edd} = 4\pi G \Mbh m_p c/\sigma_{\rm T}$ where $m_p$ is the proton mass, $\sigma_{\rm T}$ is the Thompson cross section, $c$ is the speed of light, so that $\dot{\Mbh} = \min\left(\dot{M}_{\rm BHL}, L_{\rm Edd}/(\epsilon_r c^2)\right)$.

\begin{figure*}
  \includegraphics[width=.9\linewidth]{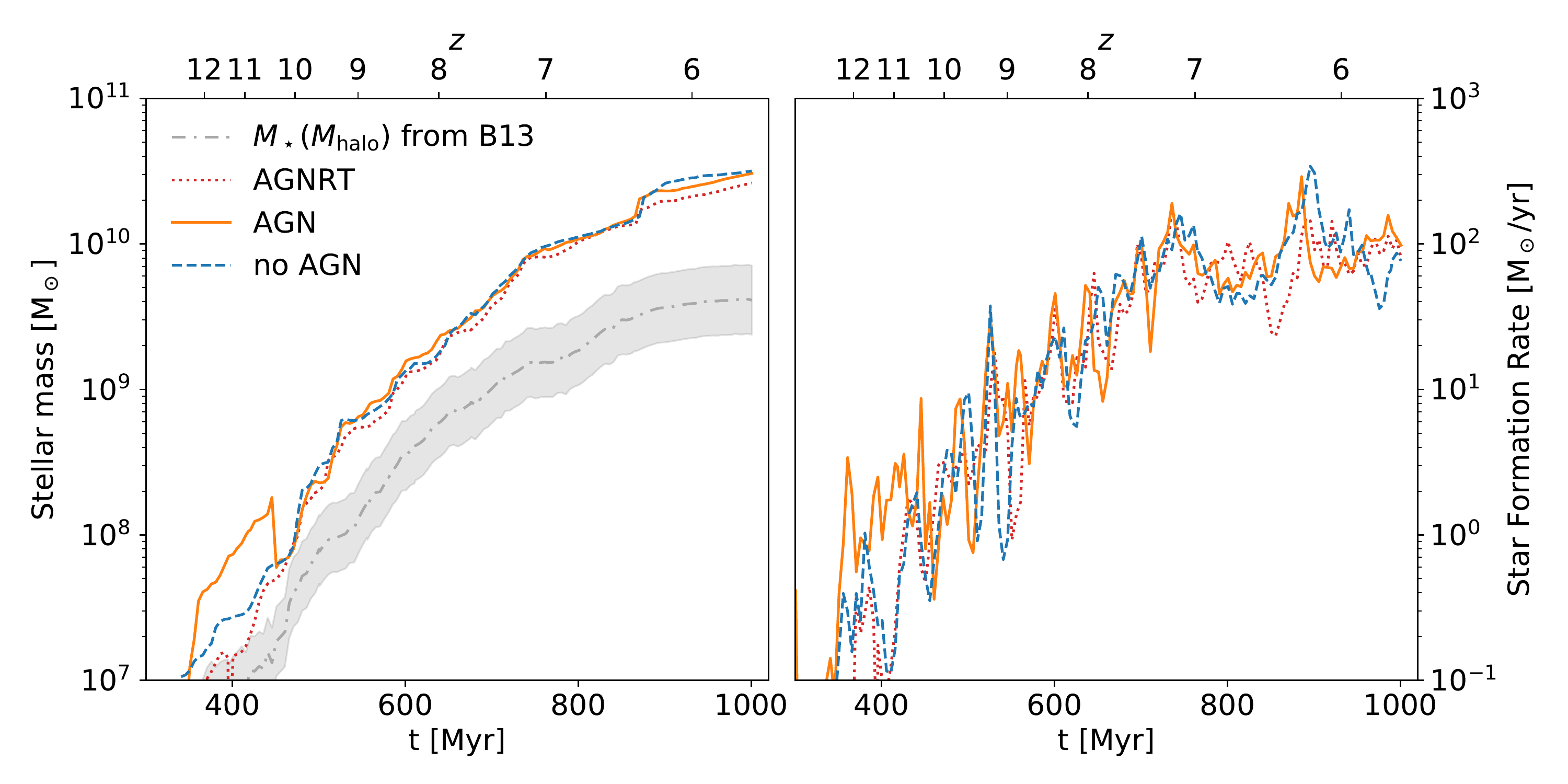}
  \caption{Star formation history of the main galaxy in our simulation for each run: dotted red line for the AGNRT run, an solid orange line for the AGN run, and a dashed blue line for the run without AGN. The stellar mass (left) steadily grows until the galaxy reaches $\sim \SI{e10}{\Msun}$, with a major merger around $z\sim 6.3$. In the meantime, the SFR (right) follows a similar trend, evolving from very variable at early times to a more regulated regime at $z \lesssim 7$. We show the mass growth expected from the stellar to halo mass relation from \citet{Behroozi2013} only as a guide to display the typical expectation from abundance matching type models (see text for details).}
  \label{fig:mstar_vs_time}
\end{figure*}

The accretion onto a BH results in AGN feedback, modelled here using the dual mode implementation of \citet{Dubois2012}: at low Eddington ratio $\lambda_{\rm Edd} = \dot{M}_{\rm BHL} / \left(L_{\rm Edd}/(\epsilon_r c^2)\right) < 0.01$, the AGN is in ``radio mode'', and in ``quasar mode'' when $\lambda_{\rm Edd} \geq 0.01$. The details of the feedback implementation are given in \citet{Trebitsch2019}, but we sketch here the main elements of the model. For both feedback modes, the AGN injects energy at a rate $\dot{E}_{\rm AGN} = \epsilon_f \epsilon_r \dot{\Mbh} c^2$, proportional to the accretion rate $\dot{\Mbh}$.
Quasar mode feedback is modelled by releasing purely thermal energy in a sphere of radius $\Delta x$ centred on the BH with a coupling efficiency $\epsilon_f = 0.15$,
For the radio mode, we deposit energy and momentum as a bipolar outflow aligned with the total angular momentum of the accreted gas with a coupling efficiency is assumed to be $\epsilon_f = 1$. The jet velocity\footnote{While this velocity is high than the reduced speed of light, we have checked that this does not affect our results, for two main reasons. First, as the jet propagates, it will very quickly decelerate below $\tilde{c}$. Second, as discussed in \citet{Trebitsch2019}, the BH spends most of its lifetime at $z\lesssim 7.5$ in a high accretion state. As a result, only a very small fraction of the timesteps are affected by this.} is fixed to be \SI{e4}{\km\per\second} with a mass loading factor of the jet 100.
The feedback efficiencies $\epsilon_f$ in both the radio and quasar modes have been empirically determined in \citet{Dubois2012} in order to reproduce the BH-to-bulge mass relations at $z=0$.

Finally, we take particular care of the detailed dynamics of the BH in this simulation. Indeed, given the fairly low mass of our BH seed compared e.g. to the DM mass resolution, we need to ensure that the dynamical friction force on the BH is taken into account below the grid \citep[see e.g.][]{Tremmel2015, Pfister2017}. For this, we follow the approach of \citet{Pfister2019} to model the dynamical friction exerted both by the gas and by the collisionless particles (stars and DM), which we do not resolve directly in our simulation. For the dynamical friction exerted by the gas, the matter lagging behind the BH induces a drag force \citep{Ostriker1999}, that we model following \citet{Dubois2013}. This frictional force is proportional to $F_{\rm DF} = \alpha f_{\rm gas} 4\pi \rho (G \Mbh / \bar{c_s}^2)$, with $\alpha  = (\rho/\rho_{\rm DF, th})^2$ if $\rho > \rho_{\rm th}$ and 1 otherwise is an artificial boost, and $f_{\rm gas}$ is a fudge factor varying between 0 and 2 and which depends on the BH Mach number, given by the ratio of the relative velocity between the BH and the gas $\bar{v}_{\rm rel}$ and the sound speed $c_s$, $\mathcal{M}_\bullet = \bar{v}_{\rm rel}/\bar{c}_s$ \citep[e.g.][]{Chapon2013}. In this work, we take $\rho_{\rm DF, th} = 50\,\mbox{cm}^{-3}$.
For the dynamical friction caused by the collisionless particles (stars and DM), we use the implementation of \citet{Pfister2019}: the (negative) acceleration of the gas is again  caused by matter lagging behind the BH, and is a function of the BH mass, velocity, and of the detailed distribution of stars and DM within $4\Delta x$ of the BH. We note that the implementation for collisionless particles is similar to that of \citet{Tremmel2015}.

\subsection{Gas cooling and heating}
\label{sec:sims:cooling}
\ramses features non-equilibrium cooling for hydrogen and helium by tracking the abundances of H, H$^+$, He, He$^+$, He$^{++}$, as well as metal cooling implemented by a set of tabulated cooling rates computed with \textsc{Cloudy}\footnote{\url{http://www.nublado.org/}} {\citep[last described in][]{Ferland2017}} above \SI{e4}{\K}. Below \SI{e4}{\K}, we account for energy losses via metal line cooling following \citet{Rosen1995} and scaling the metal cooling enhancement linearly with the gas metallicity, assuming solar abundance pattern for the metals. We currently do not take into account the impact of the local ionizing flux on metal cooling, but instead assume photo-ionization equilibrium with a redshift dependent \citet{Haardt1996} UV background for the metals. We stress that this UV background is not used for the hydrogen and helium non equilibrium photo-chemistry, for which we use the local photon field transported self-consistently by the RT solver.

\section{Results}
\label{sec:results}

Throughout this work, we will discuss three flavours of our simulation: our fiducial run (``AGNRT''), which includes all the physics described in Sect.~\ref{sec:method}; a run where the feedback from the AGN is purely thermal/mechanical (as in e.g. \citealt{Dubois2012}) and where the radiation is only produced by stellar populations (``AGN''); and a third control run where we include no BH at all (``no AGN'').
Fig.~\ref{fig:all_galaxies} presents a face-on view of the main galaxy of the simulation for each run (AGNRT, AGN and with no AGN, from left to right), showing the distribution of stars on the upper row and SFR surface density on the lower row. There is no significant difference in terms of size or morphology between each run.

\subsection{A growing black hole in a growing galaxy}
\label{sec:BH-galaxy-coevolution}
We illustrate in Fig.~\ref{fig:mstar_vs_time} the assembly of the main galaxy as a dotted red line for the AGNRT run, a solid orange line for the AGN run, and a dashed blue line for the run without AGN, and we will keep this colour coding for all other figures unless specified otherwise.
In all three simulations, the main galaxy (illustrated at $z\sim 5.7$ in Fig.~\ref{fig:all_galaxies}) grows steadily from the dwarf regime ($\Mstar \sim \SI{e7}{\Msun}$) at $z\sim 12$ to a mass of $\Mstar\sim \SI{3e10}{\Msun}$ by $z \sim 6$.
The sudden increase followed by a drop of the stellar mass in the `AGN' simulation is due to a mis-identification of the galaxy in the merger tree at early times.
We also show on the left panel the range of stellar masses expected from the model of \citet{Behroozi2013} give the growth of the main halo in our simulation: this gives a qualitative idea of how fast our galaxy is expected to grow given its host halo growth (we checked that others models, such as those of \citealt{Moster2018,Behroozi2019} give a similar growth). In all runs, the galaxy appears over-massive compared to its halo: we note however that abundance matching type techniques are still highly uncertain at high redshift, especially in the low mass regime \citep[e.g.][]{Moster2018,Behroozi2019}. At the end of our simulation, when the galaxy lives in a relatively massive halo, the stellar mass still appears higher than empirical models suggest: this suggests either that the star formation is not strongly enough regulated (e.g. by stellar feedback) in our simulation, or that the comparison of our stellar mass to observations is too indirect, or a combination of the two factors. Regarding this last point, \citet{Behroozi2019} highlight that at $z\gtrsim 4$, the constraints on galaxy growth are predominantly relying on converting the UV luminosity to a stellar mass, which at very high redshift can be severely uncertain. Nevertheless, we still address the possibility that our stellar feedback might not be efficient enough at limiting the star formation in the galaxy. \citet{Rosdahl2018} have shown that while the strength of the SN feedback directly affects the stellar-to-halo mass relation, it plays very little role in the reionization history of their simulation.

\begin{figure}
  \includegraphics[width=\columnwidth]{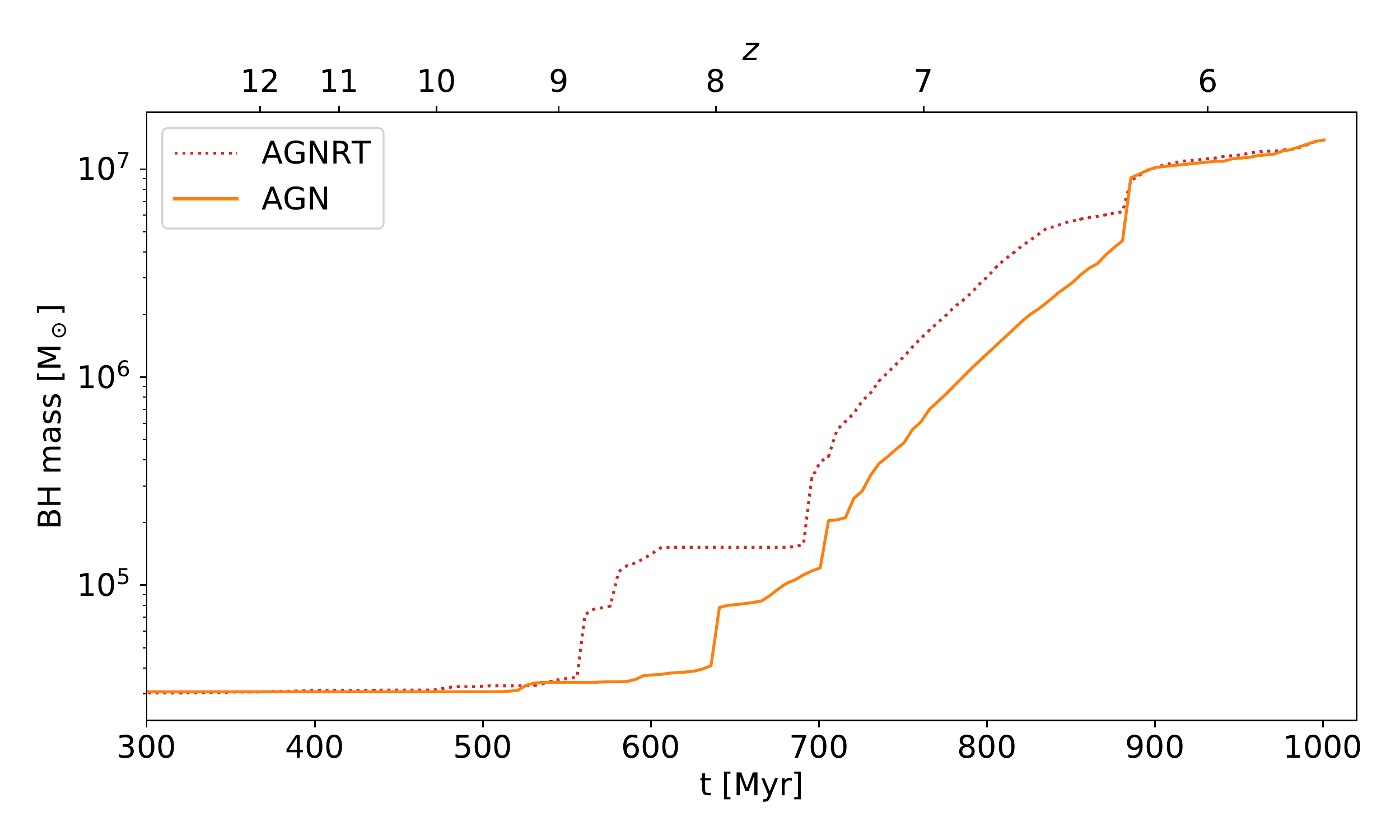}
  \caption{BH growth history for the central BH of the main galaxy in the AGNRT and AGN runs. In both cases, the BH growth starts around the time the galaxy reaches $\Mstar \sim \SI{e9}{\Msun}$, accompanied by a few BH-BH mergers, and then grows rapidly to $\Mbh \sim \SI{e7}{\Msun}$ by $z \sim 6$}
  \label{fig:BH_growth}
\end{figure}

The star formation rate (SFR, right panel) evolution can be split in two epochs: first, at $z > 7.5$, the SFR increases quickly from below $\SI{1}{\Msun\per\year}$ up to around \SI{100}{\Msun\per\year} when the galaxy reaches $\Mstar \sim \SI{e10}{\Msun}$. After that, the galaxy reaches some form of self-regulation and the SFR remains constant with some fluctuations around this value.
This behaviour is common to all three simulations: this strongly suggests that the feedback from the AGN is not playing a major role in setting the star formation properties of the host galaxy. This happens in spite of the fact that the central BH is actively growing, as shown in Fig.~\ref{fig:BH_growth}: after an initial phase where the BH is not growing due to the strong SN feedback preventing gas from settling in the vicinity of the BH \citep[e.g.][]{Dubois2015}, accretion onto the BH becomes very efficient (with Eddington ratio $\lambda_{\rm Edd} \simeq 10\% - 100\%$) after the galaxy has reached $\Mstar \gtrsim \SI{e9}{\Msun}$, leading to a mass of $\Mbh \gtrsim \SI{e7}{\Msun}$ by $z \sim 6$. This transition happens at a similar stellar mass than found by previous studies \citep[e.g.][]{Dubois2015,Habouzit2017}.
Combining this with the results of \citet{Trebitsch2019}, this points toward a picture where the galaxy and the SMBH are regulating their own mass growth once the galaxy is massive enough, independently of one another (albeit ultimately feeding from the same gas reservoir).

\begin{figure}
  \includegraphics[width=\columnwidth]{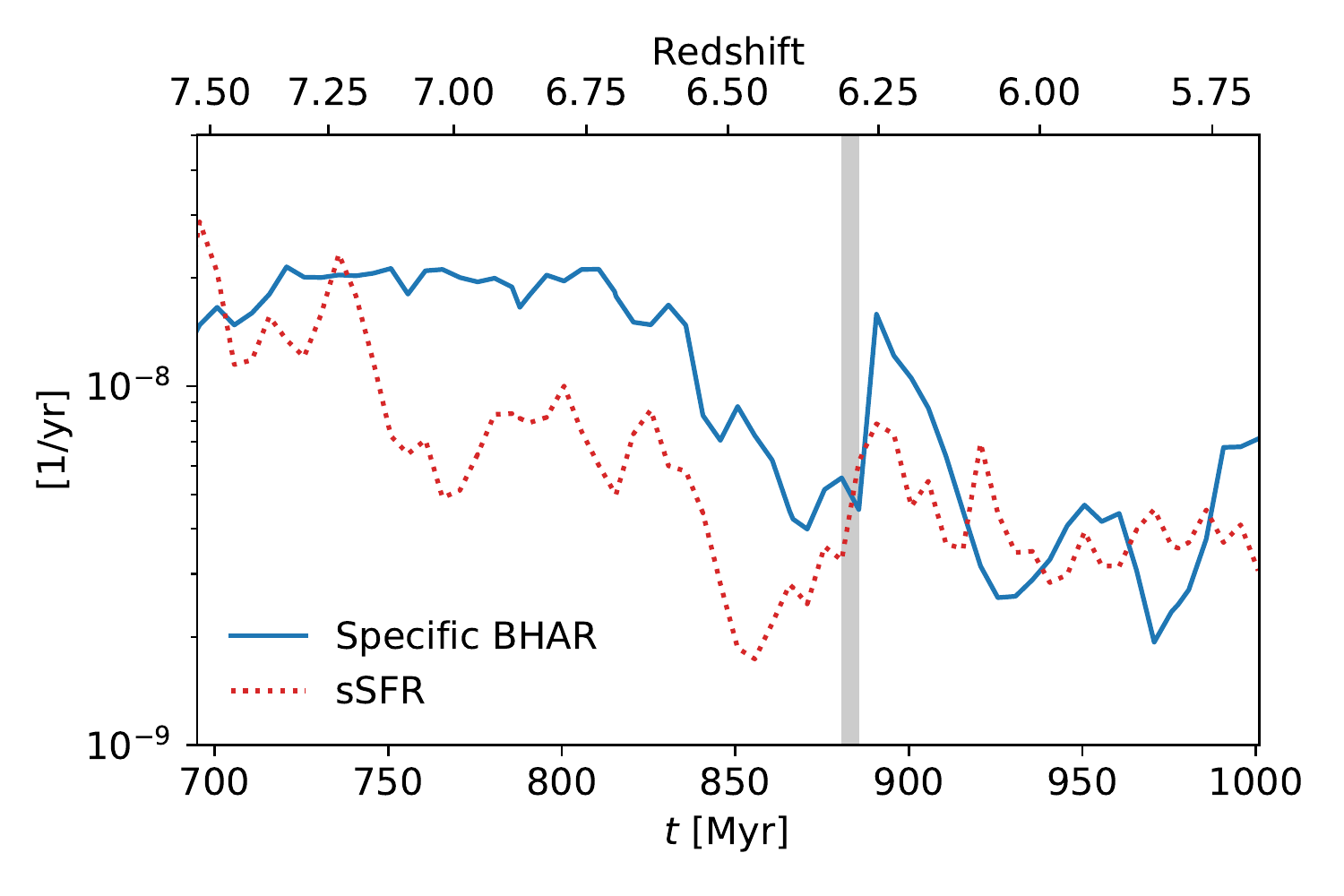}
  \caption{Specific SFR (in red) and BHAR (in blue) for the AGNRT run around the time of the last galaxy merger, around $z \sim 6.3$. While the BH growth is increased, the duration of the boost is small ($\lesssim 50$ Myr).}
  \label{fig:specific_rates_merger}
\end{figure}
Fig.~\ref{fig:mstar_vs_time} and Fig.~\ref{fig:BH_growth} show a distinctive feature around $t \sim \SI{900}{\mega\year}$: a sudden jump in stellar mass and BH mass accompanied by a sharp increase of the SFR. This corresponds to a major merger with mass ration $\sim 1\!:\!4$, followed by a BH-BH merger. This merger has a small but noticeable effect on the BH accretion rate (BHAR) and to some extent on the SFR, but the effect dissipates quickly. This is shown in Fig.~\ref{fig:specific_rates_merger}: right after the two galaxy merge into one (vertical grey line), both the specific SFR and the specific BHAR reach a peak (although this is less significant for the specific SFR).

\subsection{Escape of ionizing radiation}
\label{sec:effect-agn-escape}

We now turn our attention to the ionizing output of our simulated galaxies, with the goal of better quantifying the contribution of bright $\sim L^\star$ galaxies to the reionization.

\subsubsection{Contribution from stellar populations}
\label{sec:fesc:stars}

\begin{figure}
  \includegraphics[width=\columnwidth]{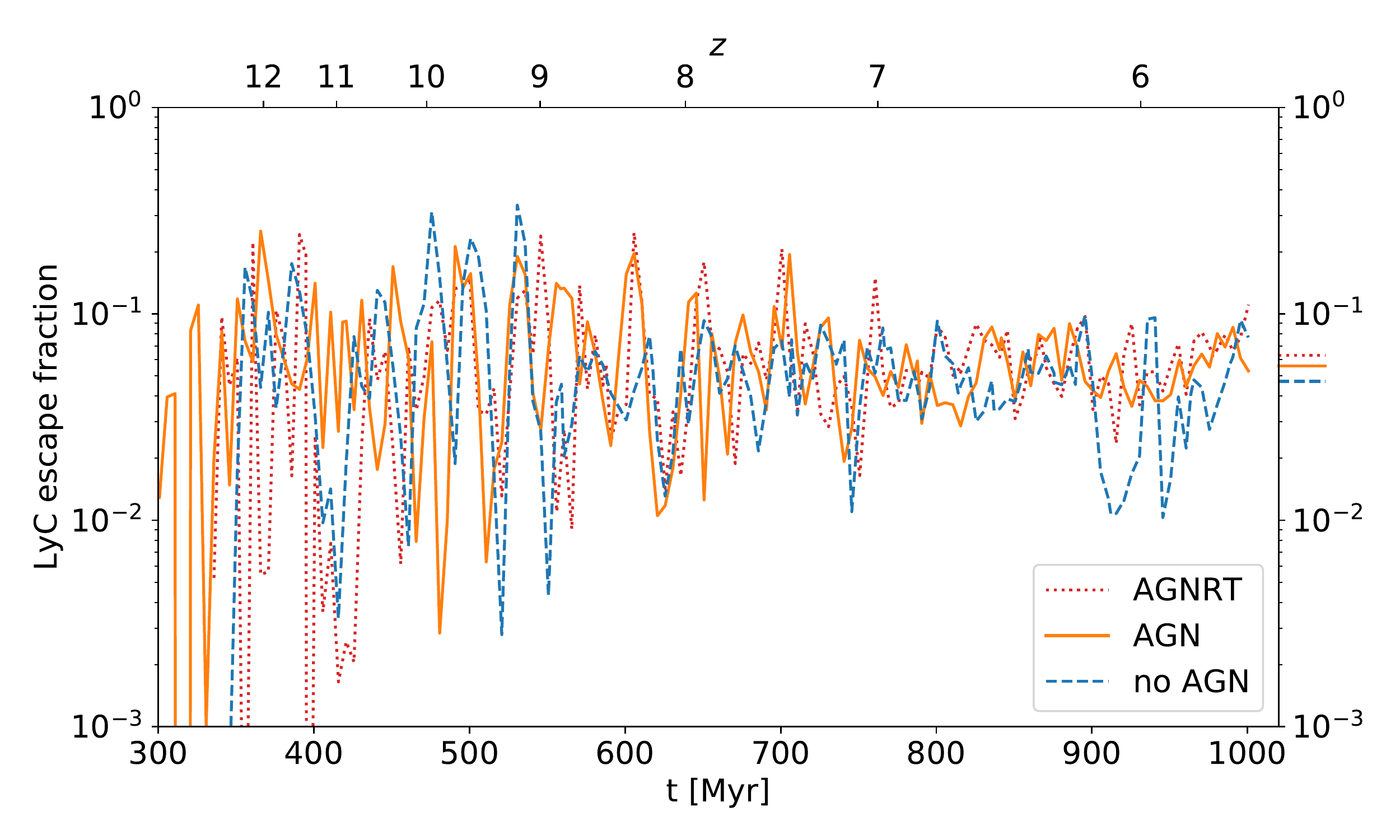}
  \caption{Ionizing escape fraction measured at the virial radius for the three runs. For all runs, \fesc varies quickly, and the amplitude of the variations decreases as the galaxy settles, finally reaching $\fesc \sim 5\%$}
  \label{fig:fesc_vs_time}
\end{figure}

We quantify the amount of ionizing radiation produced by the stellar populations in our galaxy and escaping into the IGM by measuring the luminosity-averaged escape fraction \fesc:
\begin{equation}
  \label{eq:fescray}
  \fesc = \frac{\sum_i L_{\rm ion}^i \bar{T}_i}{\sum_i L_{\rm ion}^i},
\end{equation}
with $\bar{T}_i = \langle e^{-\tau_{\hi}^{i,j}} \rangle_{j}$ the angle-averaged transmission for the $i^{\rm th}$ star particle, and $L_{\rm ion}^i$ its ionizing luminosity\footnote{\label{fn:fesc}As we perform RHD simulations, we could in principle measure the ratio of the total ionizing flux crossing the virial radius divided by the intrinsic ionizing production, as e.g. in \citet{Kimm2014, Trebitsch2017}. This however does not work when galaxies regularly experience mergers, as we discuss in detail in Appendix~\ref{sec:app:fesc}.}.
We measure this quantity for our three runs by casting rays from each star particle within $0.3 \Rvir$ using the \textsc{Rascas} code (Michel-Dansac et al., submitted) and present the results in Fig.~\ref{fig:fesc_vs_time} with the same colour-coding as in Fig.~\ref{fig:mstar_vs_time}. The three dashes on the right axis indicate the average escape fraction measured after $t > \SI{750}{\mega\year}$, when the SFR remains constant.
On average, the three runs present a very similar behaviour, with a fairly low average escape fraction of $\fesc \sim 5-7\%$, consistent with the recent results of \citet{Steidel2018} on a sample of LBGs at $z\sim 3$ and with the detailed simulation of \citet{Yoo2020} of an isolated galaxy of similar mass.
A striking feature of Fig.~\ref{fig:fesc_vs_time} is that as the galaxy grows, the variability in \fesc decreases strongly: at $z\sim 10$, when the stellar mass of the galaxy is around $\Mstar \sim \SI{e8}{\Msun}$, the escape fraction can vary by up to two orders of magnitude in $\sim \SI{10}{\mega\year}$, while the fluctuations become milder at $z \lesssim 7$, when the galaxy reaches a more regulated state with $\Mstar \gtrsim \SI{e10}{\Msun}$. This is expected from the picture in which feedback processes associated to star formation create channels through which radiation can escape \citep[e.g.][]{Wise2009, Kimm2014,Trebitsch2017}. Indeed, as star formation is extremely bursty in low mass systems, the number of simultaneously star forming regions is low, so that one Lyman-leaking channel is enough for a large fraction of the ionizing radiation produced to escape: the galaxy is either ``on'' or ``off''. When the galaxies are more massive, this is no longer true, and the galaxy-averaged escape fraction will be lowered by the large number of star forming regions embedded in dense \hi clouds. These results are consistent with the model of \citet{Howard2018}, who estimate the escape fraction of synthetic galaxies by averaging over populations of star forming clouds, finding that their dwarf models systematically yield higher average \fesc and stronger fluctuations than their models for spiral galaxies.

\begin{figure}
  \includegraphics[width=\columnwidth]{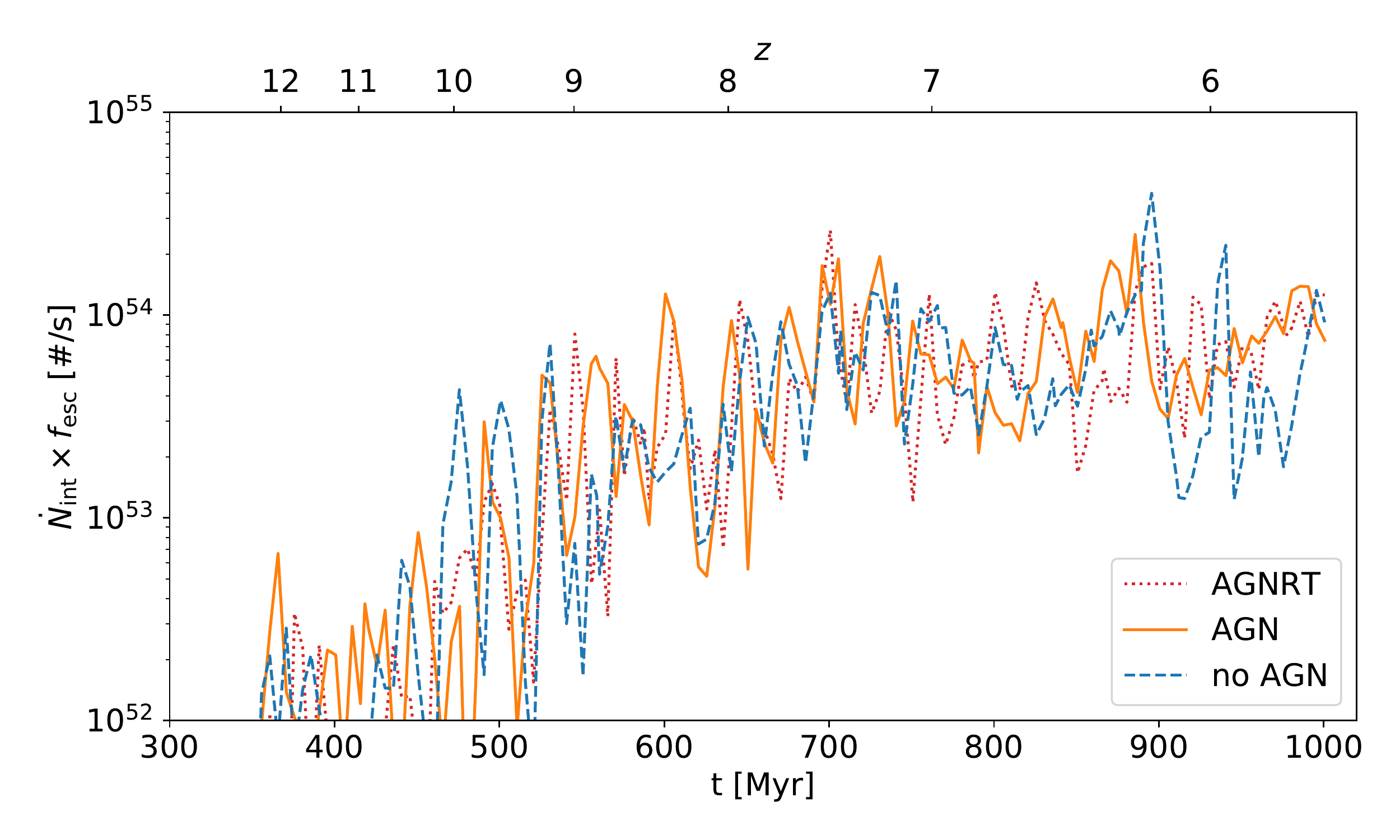}
  \caption{Evolution of the ionizing flux escaping the main halo ($\dot{N}_{\rm esc} = \dot{N}_{\rm int} \fesc$) for the three runs.}
  \label{fig:Nesc}
\end{figure}
We can now estimate the ionizing luminosity of our simulated galaxy in each run as $\dot{N}_{\rm esc} = \dot{N}_{\rm int} \fesc$, where $\dot{N}_{\rm int} = \sum_i L_{\rm ion}^i$ is the total intrinsic ionizing luminosity of the galaxy. We summarize this in Fig.~\ref{fig:Nesc}, keeping the same colour-coding as before. Apart from the rapid fluctuations due to the quickly varying \fesc, the evolution of the escaping flux $\dot{N}_{\rm esc}$ broadly follows that of the SFR: it rises until $z \sim 7.5$, and stay roughly constants after that. This behaviour is the same for all three runs, suggesting again that the AGN is not strongly affecting the gas distribution in and around star forming regions.
Interestingly, we do not see a very clear sign of the major merger that occurs around $z \sim 6.3$ in the evolution of \fesc, and only a marginal trend in the evolution of $\dot{N}_{\rm esc}$. This is partly due to the already important variations in \fesc, and further confirms that the escape and production of ionizing radiation is a process very local to star forming clouds.

\subsubsection{Contribution of the AGN to the LyC leakage}
\label{sec:fesc:agn}

\begin{figure}
  \includegraphics[width=\columnwidth]{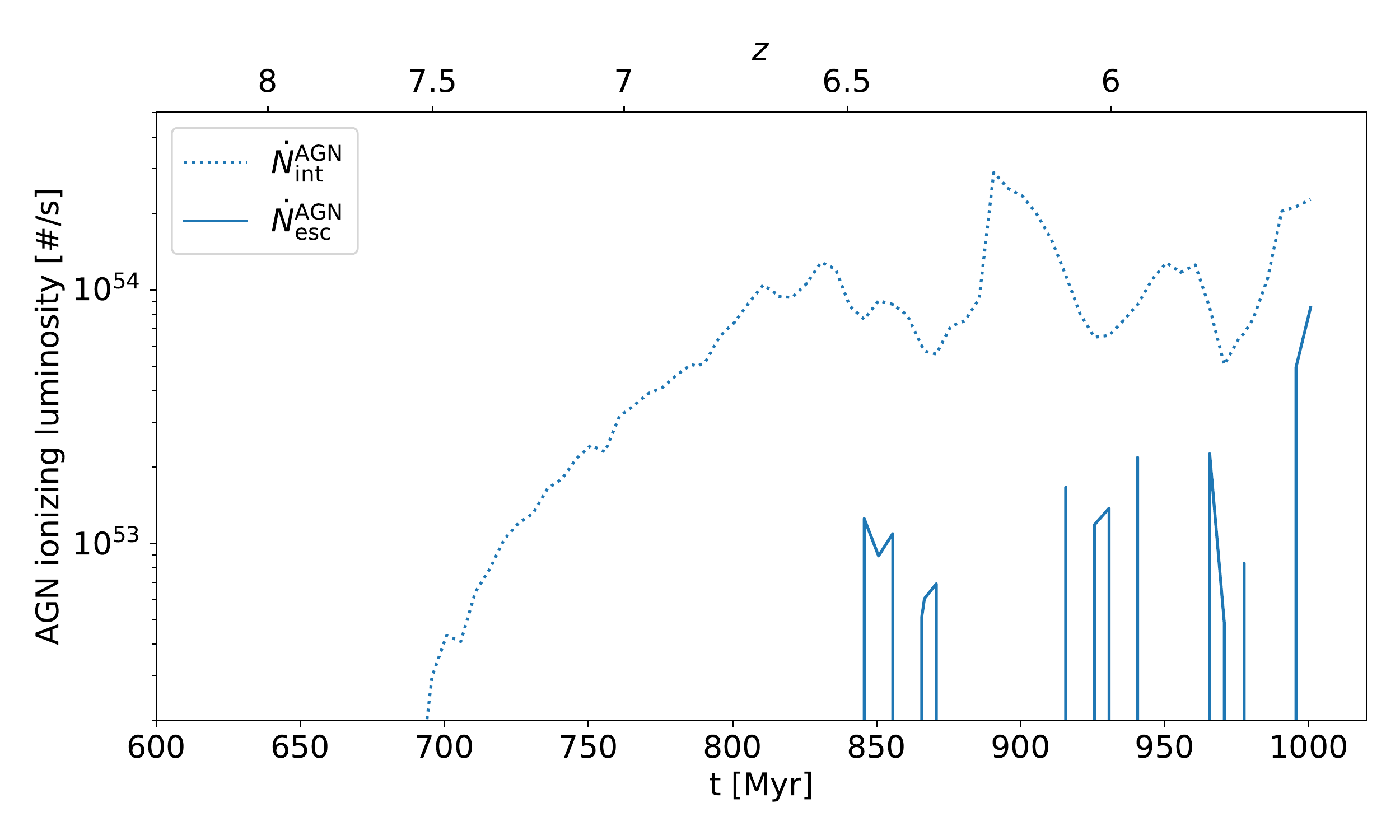}
  \caption{Comparison of the intrinsic (dotted line) and escaping (solid line) luminosity produced by the AGN in the AGNRT run. Only a tiny fraction of the ionizing photons produced can escape the halo, because most of the time the AGN is obscured.}
  \label{fig:NdotAGN}
\end{figure}

As mentioned previously, the central BH in the main galaxy is actively growing, both in the AGN and AGNRT runs, suggesting that the AGN is an important source of ionizing radiation: this can be seen as the dotted line in Fig.~\ref{fig:NdotAGN}, illustrating the \hi-ionizing radiation produced by the AGN in the AGNRT run\footnote{Note that the axes on Fig.~\ref{fig:Nesc} and Fig.~\ref{fig:NdotAGN} are different.}. This ionizing luminosity is lower than that of stellar populations by approximately one order of magnitude, but slightly higher than their escaped ionizing luminosity.
When looking at the flux escaping the halo (solid line in Fig.~\ref{fig:NdotAGN}), we find a non-zero $\dot{N}_{\rm esc}^{AGN}$ only a small fraction of the time. In other words, the total escaping luminosity coming from the AGN is negligible most of the time.

This might seem at odds e.g. with the study of \citet{Grazian2018}, who found that for their sample of AGN, the average escape fraction is of the order of $\sim 75\%$, thus extending the earlier work of \citet{Cristiani2016} on bright quasars. Similarly, \citet{Guaita2016} detected LyC flux with a relative escape fraction $\fescAGN_{\rm rel} \sim 0.72 \pm 0.18$ for one object with $\Magn \sim -21.9$ at $z\sim 3.46$, but they could only put upper limits on their other seven AGN.

A closer look at the selection criterion of \citet{Grazian2018} can however largely explain this apparent discrepancy. As all the AGN selected in their sample have $-25 \lesssim \Magn \lesssim -23$, they are not strongly obscured.
Contrasting to this, we have shown in \citet{Trebitsch2019} that the AGN in our simulation is most of the time surrounded by a column density of \hi in excess of $\NH > \SI{e20}{\per\cm\squared}$, corresponding to an optical depth $\tau \gg 100$ for the ionizing radiation. This large column of \hi is therefore enough to completely absorb all ionizing radiation produced by the AGN, except for rare episodes (corresponding to the spikes of $\dot{N}_{\rm esc}^{AGN}$ seen in Fig.~\ref{fig:NdotAGN}). This is qualitatively consistent with the results of e.g. \citet{Cowie2009}, who found that only their quasars displaying broad emission lines are seen in the ionizing UV, and that the ionizing luminosity of the rest of their sample is consistent with zero. 
We note that even if our estimate of the nuclear obscuration is uncertain, there is a large amount of neutral gas in the ISM of the galaxy that contributes to lowering \fescAGN. This is well in line with the results of \citet{Circosta2019}, who found that for a sample of bright $z > 2.5$ quasars, the ISM of the host galaxy strongly contributes to the total obscuration.

Finally, our findings are qualitatively consistent with the observations of \citet{Micheva2017}, who found strong evidence for a low \fescAGN from a sample at $z \sim 3$. Interestingly (but keeping in mind that it is hard to compare one simulated galaxy to a single observed one), they assess that for one of their AGN, the detected LyC flux is dominated by the stellar populations in the galaxy: this is exactly how our simulated galaxy would be classified.

\subsection{Bright galaxy, faint AGN}
\label{sec:galaxy-agn-lumin}

We have shown that while the main galaxy in our simulation hosts an actively growing BH (at least for the runs with BH), the LyC flux is completely dominated by the stellar populations. We will now extend this analysis to other wavelengths, in order to determine if the object we are focusing on should rightfully be called a ``star forming galaxy'' or an AGN.
We stress that we do not focus on any emission line properties (neither metal lines nor hydrogen recombination lines) in this study, which would require a more careful treatment, like post-processing the simulation with \textsc{Cloudy} as in \citet{Hirschmann2017}. We therefore cannot determine where our galaxy would lie in various diagnostic diagrams, and only focus on the continuum emission.
This way, we follow the approach of \citet{Volonteri2017} and compare the relative contribution of the stellar populations and the AGN to the rest-frame UV and hard X-ray luminosity of the galaxy.
This is particularly relevant in the context of understanding the nature of the X-ray selected sources observed by \citet{Giallongo2015}, as both the \SIrange{2}{10}{\kilo\eV} luminosity and the UV magnitude \Magn of our simulated AGN are close to the typical values of their sample.
In the following, we will only focus on the AGNRT run, which is well justified since we have shown that the global properties of the galaxy and AGN in all three runs are comparable.

\subsubsection{Hard X-rays}
\label{sec:hard-x-rays}

\begin{figure}
  \centering
  \includegraphics[width=\columnwidth]{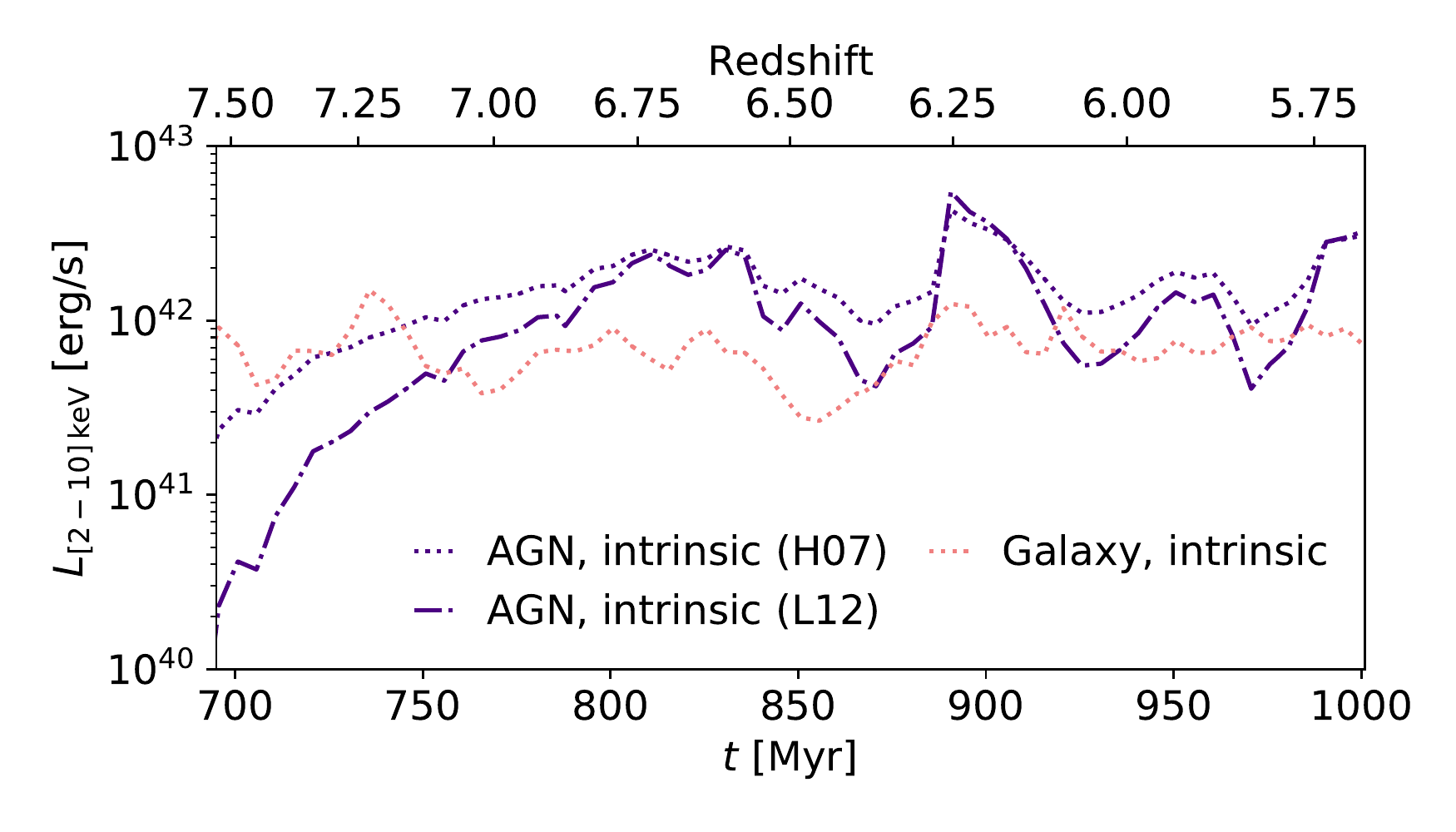}
  \caption{Intrinsic X-ray luminosity of the AGN (in purple) and of the stellar populations in the galaxy (in red). Overall, the AGN dominates the X-ray emission of the system.}
  \label{fig:X_galaxy_agn}
\end{figure}

We start by estimating the hard X-ray luminosity of our simulated galaxy in the \SIrange{2}{10}{\kilo\eV} band and the relative contribution of the AGN and stellar populations, displayed in Fig.~\ref{fig:X_galaxy_agn} (purple for the AGN, red for the galaxy).
For the stellar population, we follow \citet{Fragos2013} to estimate the X-ray luminosity of X-ray binaries (their model 245) as $L_X \propto \alpha\, (\Mstar/\SI{e10}{\Msun}) + \beta\, {\rm SFR}$, and we use the bolometric correction from \citet{Hopkins2007} (H07, dotted line) for the AGN X-ray luminosity.
We also display in Fig.~\ref{fig:X_galaxy_agn} the AGN hard X-ray luminosity estimated using the bolometric correction of \citet{Lusso2012} for Type 1 AGN (L12, dash-dotted line).
Here, contrary to the UV, the luminosity is dominated by the AGN, except at very early times ($t \ll \SI{750}{\mega\year}$) when the BH is still not very massive.
Overall, the total X-ray luminosity is comparable (or even a bit lower) to that of the systems probed by \citet{Giallongo2015}, suggesting that the X-ray emission in these objects is indeed powered by nuclear activity.

Comparing to \emph{Himiko} as a prototypical bright galaxy in the reionization era, we find that at all times, the total X-ray luminosity of our system is below the lower-limit coming from non-detection of X-rays in Himiko \citep{Ouchi2013}: this means that with a similar survey, no X-ray would have been detected in a galaxy like the one we are discussing in this paper. We note that \citet{Baek2013} have predicted that, based on its \lya properties, \emph{Himiko} should not host an AGN: they however assume for this that the \lya is powered by the AGN, while we have no evidence for this at all in our simulation. Further studies are required to make any statement on the \lya emission and observability for our galaxy, but the large amount of dust surrounding the AGN and the fact that it does not dominate the ionizing budget of the system suggest that it would not be dominating the \lya output.

\subsubsection{UV properties}
\label{sec:uv-properties}

\begin{figure}
  \centering
  \includegraphics[width=\columnwidth]{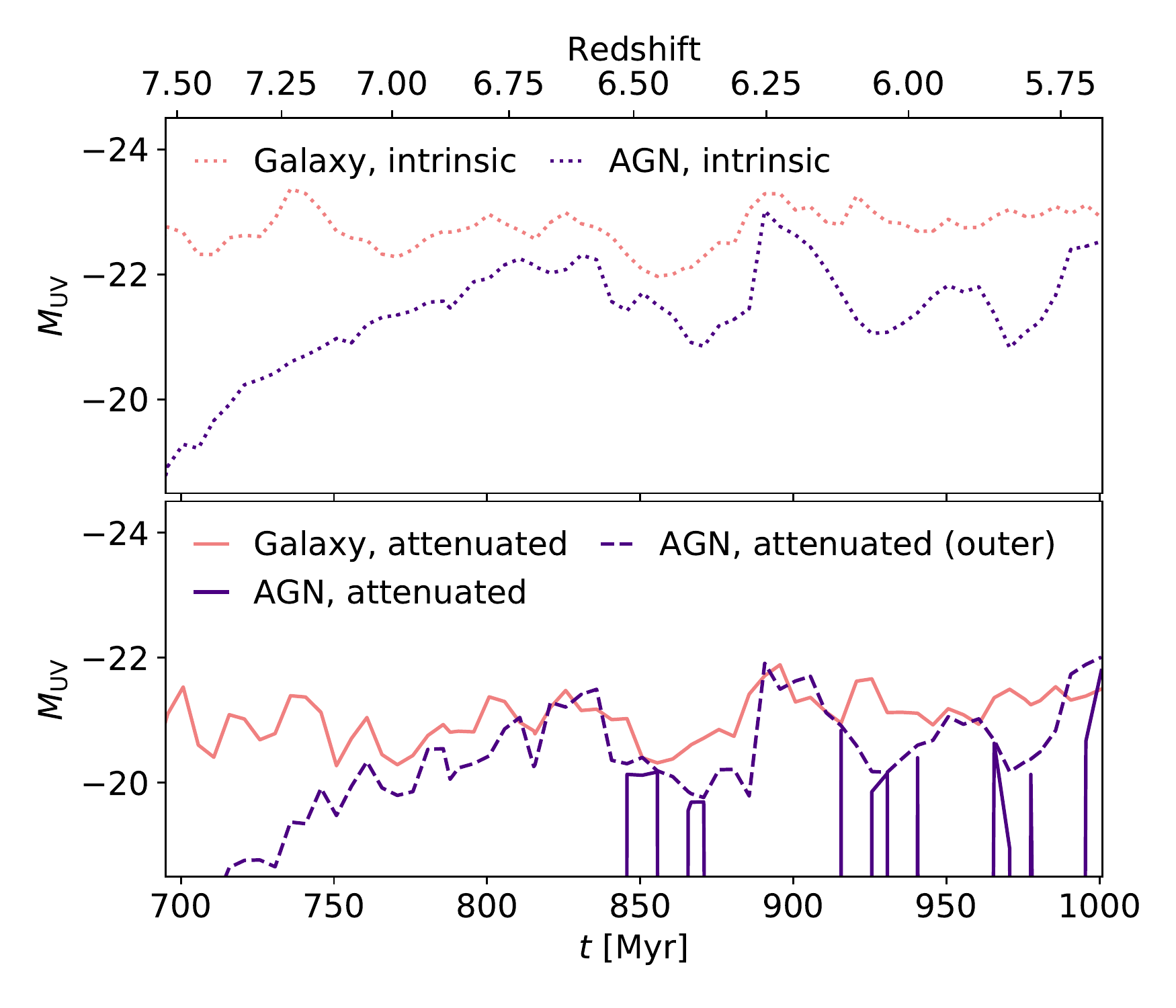}
  \caption{UV magnitude after $z=7.5$ for the galaxy (in red) and the AGN (in purple) in the main simulation of this study. The intrinsic emission is shown with dotted lines (upper panel), and the solid lines show the rest-frame UV magnitude after dust attenuation in the ISM (lower panel). The purple dashed line on the lower panel shows the UV magnitude of the AGN removing the attenuation in the inner 40 pc. Overall, the galaxy dominates the UV luminosity of the system.}
  \label{fig:muv_galaxy_agn}
\end{figure}
We now turn to the (non-ionizing) UV properties of our simulated object, as illustrated in Fig.~\ref{fig:muv_galaxy_agn}: the purple (red) lines show the AGN (galaxy) rest-frame UV magnitude, with the dotted lines indicating the intrinsic emission and the solid lines taking into account the attenuation by dust.
Additionally, the purple dashed line correspond to the AGN UV magnitude attenuated only by the ISM dust, not taking into account the innermost 40 pc surrounding the BH: this is effectively an upper limit on the AGN UV luminosity.

For the galaxy, the UV luminosity is derived directly from the properties of the stellar populations using the \citet{Bruzual2003} stellar population synthesis model, and for the AGN we convert the bolometric luminosity in UV magnitude assuming the bolometric correction of \citet{Runnoe2012}.
Both for the AGN and the galaxy, we take into account the dust obscuration as in \citet{Trebitsch2019}: we use again the \textsc{Rascas} tool \citep{MichelDansac2020} to cast rays from each star particle\footnote{For the AGN, we directly use \textsc{Yt} to cast rays from the sink particle.} in the simulation and integrate the dust optical depth along each ray as
\begin{equation}
  \label{eq:taudust}
  \tau_d(\lambda) = \int_{\rm ray} n_d(\ell) \sigma_d(\lambda) d\ell,
\end{equation}
where $\sigma_d(\lambda)$ is the dust interaction cross section per hydrogen atom defined by the fits of \citet{Gnedin2008} for their Small Magellanic Cloud (SMC) model and $n_d$ a pseudo-number density of dust grains, given by $n_d = (n_{\hi} + f_{\rm ion} n_{\hii})Z/Z_0$ \citep{Laursen2009}, with $Z_0 = 0.005$ is the mean metallicity of the SMC. Following \citet{Laursen2009}, we take $f_{\rm ion} \sim 0.01$ as the typical dust to gas ratio in ionized gas.
The stellar populations dominate the UV production of the galaxy most of the time, either prior or after dust attenuation: this is qualitatively consistent with our conclusions regarding the ionizing UV production. Even if we discard the circumnuclear region in our simulation to compute $\tau_d$ (dashed purple line), the AGN never really dominates the UV budget of the galaxy.

\begin{figure}
  \centering
  \includegraphics[width=\columnwidth]{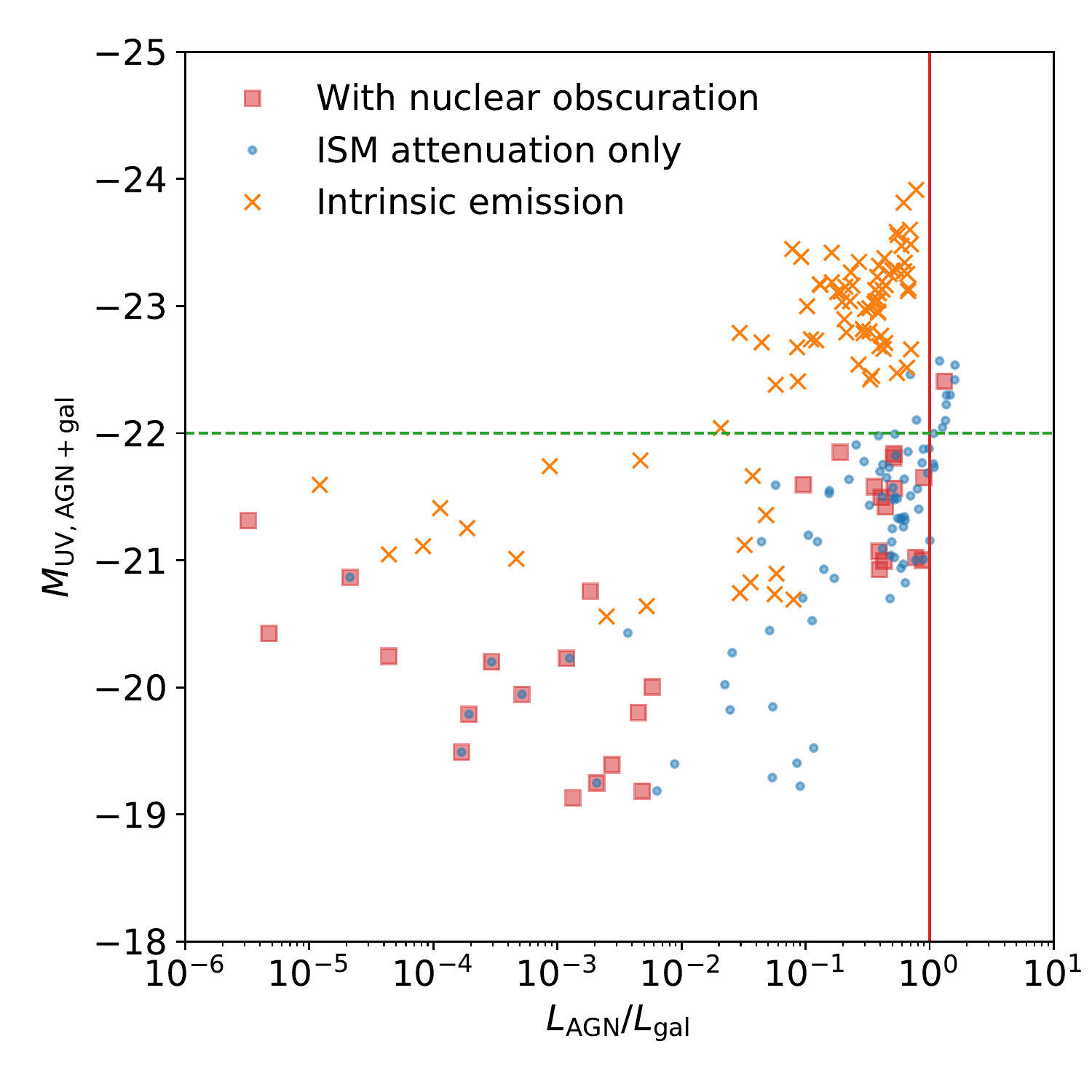}
  \caption{Evolution of the total UV luminosity of the system as a function of the ratio of the contributions from the AGN and the stellar populations taking into account different levels of AGN attenuation (orange crosses for no obscuration, blue dots for ISM obscuration, and red squares for ISM + nuclear obscuration). Each marker correspond to a distinct timestep of the simulation. The vertical red line indicating a similar luminosity for the AGN and the galaxy. When obscuration is taken into account, only when the total $\MUV \lesssim -22$ does the AGN dominates in the UV.}
  \label{fig:UVratios}
\end{figure}

The total UV magnitude of our object, around $\MUV \sim -22$, falls exactly within the range where \citet{Volonteri2017} predicts that the AGN UV luminosity should be at most of the order of the galaxy UV luminosity: we illustrate this in Fig.~\ref{fig:UVratios}, where we show the ratio of the AGN to galaxy luminosity, $L_{\rm AGN}/L_{\rm gal}$, as a function of the total UV magnitude of the object for successive timesteps of the simulation after $z\leq 7.5$. The vertical red line marks an equal contribution from both sources, and the horizontal green line correspond to the typical luminosity of typical $z \sim 6$ bright galaxies. The red squares and blue points correspond to the dust-attenuated UV emission including or not the circumnuclear region, respectively, while the intrinsic emission (pre-attenuation) is shown with orange crosses.
As suggested by \citet{Volonteri2017}, we find that below $\MUV \sim -22$ the AGN is always sub-dominant, and that there is a trend of increasing $L_{\rm AGN}/L_{\rm gal}$ in brighter systems.

\section{Discussion and conclusions}
\label{sec:ccl}
Interestingly, the typical value of the total UV luminosity of our object, $\MUV \sim -22$, is very close to that of the brightest \lya emitters detected at $z \gtrsim 6$ such as \emph{Himiko} or CR7 \citep[e.g.][]{Matthee2017}, and at the same time that it is just below the magnitude at which the AGN and galaxy UV LF overlap \citep[e.g.][]{Matsuoka2018, Ono2018, Stevans2018}. While the nature of bright galaxies such as CR7 is still debated, it is comforting to note that the analysis of \citet{Bowler2017} suggests it could be compatible with an (obscured) AGN, very similar to our object.

This paints a picture in which bright galaxies at $z \gtrsim 6$ are hosting actively growing black holes which are just not quite bright enough or too obscured to be dominating over the luminosity of their host. This offers a complementary insight to the results of \citet{Sobral2018}, who found that at $z \sim 2-3$, there is a sharp transition in the observed nature of bright sources around $L_{\rm UV} \sim 2 \times L^\star$, from star forming galaxies to AGN. Indeed, we suggest here that the transition is in part due to the AGN just not being dominating the UV light below that threshold, even if the BH is (almost) maximally growing.
The fact that the UV is mostly coming from young stars rather than the faint AGN in our system is not without consequences: for instance, directly inferring the AGN ionizing emissivity from the faint end of the AGN UV LF as in e.g. \citet{Giallongo2015} would significantly overestimate the contribution of AGN to the high-$z$ ionizing background. Indeed, even if $\fescAGN \sim 100\%$ (which is not what we find here), converting the UV luminosity to the ionizing band using the AGN spectral shape would not be appropriate for the (large) fraction of the UV that is actually coming from stars.

Our simulation comes however with some caveats. For instance, we have not taken into account the effect of radiation pressure (RP) from the multi-scattering of infrared radiation. Dedicated work \citep[e.g.][]{Bieri2017,Costa2018} have shown that in massive galaxies, radiation from the AGN can launch winds through this process.
  In this work, we model these winds as the `quasar mode' feedback, where winds are thermally driven.
  The simulation of \citet{Costa2018} suggests that RP-driven winds affect the ISM differently from thermally driven winds by penetrating deeper in the ISM and significantly reducing the gas density in the inner regions of the galaxy. However, they do not model the growth of the BH self-consistently in their simulation: in our case, whenever winds reduce the gas density in the vicinity of the BH, the accretion rates drops, and the BH stops being UV-bright. Additionally, we note that our quasar mode feedback efficiency is significantly higher than theirs, by a factor $\gtrsim 5$: therefore, the (thermal) energy injection in our simulation will be much higher.
  Unfortunately, we cannot directly compare the effect of RP-driven versus thermally driven winds in our simulation: \citet{Bieri2017} have shown that using the reduced speed of light approximation with $\tilde{c}$ similar to the ones we have used here can severely underestimate the mechanical advantage of the radiation-driven outflows. Nevertheless, using an isolated galaxy setup with a comparable halo and stellar mass than our target galaxy at $z \sim 6$, we found that at our resolution, the growth of the BH is not efficiently regulated by RP-driven feedback. Because of this, we need to rely on an effective description of the AGN winds, which efficiently regulates the growth of the BH \citep{Trebitsch2019}.
   Additionally, another key difference between the simulation of \citet{Costa2018} and ours is that their AGN luminosity is typically 1000 times higher than that of our BH. The effect of the quasar luminosity on the strength of the radiative feedback has been explored by \citet{Bieri2017} using very high resolution simulations. Their results indicate that for low luminosity quasars like ours, the radiation pressure driven winds do not create low-density channels through which ionizing radiation could escape.

We now summarize the main results of our study:
\begin{itemize}
\item Massive BH can grow actively in a bright LBG, but their feedback does not affect the galaxy very strongly, even at high masses ($\Mstar \gtrsim \SI{e10}{\Msun}$).
\item The ionizing output of bright LBGs is largely dominated by young stars rather than the AGN, and their typical $\fesc \sim 5\%$.
\item The feedback from the AGN does not affect the escape of ionizing radiation produced by young massive stars.
\item Deep X-ray surveys would detect the AGN in galaxies like the one we study, but the bulk of the UV luminosity (ionizing or not) would still be dominated by stellar populations.
\end{itemize}

To assess how general these conclusions are, it will be necessary to expand the number of simulated galaxies from one zoom to a large sample, which will be computationally expensive. In the meantime, the fact that our system shares many properties with bright LAEs observed at $z \sim 6$ gives a strong motivation to explore the \lya properties of our system, which requires dedicated radiative transfer modelling. We will explore both these leads in future works.

\section*{Acknowledgements}

We wish to thank the referee for an insightful report that significantly improved the manuscript.
MT thanks Harley Katz, Taysun Kimm and Joki Rosdahl for fruitful discussions and comments.
MT and MV acknowledge funding from the European Research Council under the European Community's Seventh Framework Programme (FP7/2007-2013 Grant Agreement no. 614199, project `BLACK'). For part of this work, MT acknowledges support from the Deutsche Forschungsgemeinschaft (DFG, German Research Foundation) under Germany's Excellence Strategy EXC-2181/1 - 390900948 (the Heidelberg STRUCTURES Cluster of Excellence).
This work has made use of the Horizon Cluster hosted by Institut d'Astrophysique de Paris; we thank St{\'e}phane Rouberol for running smoothly this cluster for us.
This work was granted access to the HPC resources of CINES under the allocation A0040406955 made by GENCI.
This work has made extensive use of the \textsc{Yt}\footnote{\label{fn:yt}\url{https://yt-project.org/}} analysis package \citep{Turk2011} and NASA's Astrophysics Data System, as well as the \textsc{Matplotlib} \citep{Hunter2007}, \textsc{Numpy/Scipy} \citep{Jones2001} and \textsc{IPython} \citep{Perez2007} packages.



\bibliographystyle{mnras}
\bibliography{fesc_agnrt} 

\begin{thebibliography}{}
\makeatletter
\relax
\def\mn@urlcharsother{\let\do\@makeother \do\$\do\&\do\#\do\^\do\_\do\%\do\~}
\def\mn@doi{\begingroup\mn@urlcharsother \@ifnextchar [ {\mn@doi@}
  {\mn@doi@[]}}
\def\mn@doi@[#1]#2{\def\@tempa{#1}\ifx\@tempa\@empty \href
  {http://dx.doi.org/#2} {doi:#2}\else \href {http://dx.doi.org/#2} {#1}\fi
  \endgroup}
\def\mn@eprint#1#2{\mn@eprint@#1:#2::\@nil}
\def\mn@eprint@arXiv#1{\href {http://arxiv.org/abs/#1} {{\tt arXiv:#1}}}
\def\mn@eprint@dblp#1{\href {http://dblp.uni-trier.de/rec/bibtex/#1.xml}
  {dblp:#1}}
\def\mn@eprint@#1:#2:#3:#4\@nil{\def\@tempa {#1}\def\@tempb {#2}\def\@tempc
  {#3}\ifx \@tempc \@empty \let \@tempc \@tempb \let \@tempb \@tempa \fi \ifx
  \@tempb \@empty \def\@tempb {arXiv}\fi \@ifundefined
  {mn@eprint@\@tempb}{\@tempb:\@tempc}{\expandafter \expandafter \csname
  mn@eprint@\@tempb\endcsname \expandafter{\@tempc}}}

\bibitem[\protect\citeauthoryear{{Atek}, {Richard}, {Kneib}  \&
  {Schaerer}}{{Atek} et~al.}{2018}]{Atek2018}
{Atek} H.,  {Richard} J.,  {Kneib} J.-P.,   {Schaerer} D.,  2018, \mn@doi
  [\mnras] {10.1093/mnras/sty1820}, \href
  {https://ui.adsabs.harvard.edu/abs/2018MNRAS.479.5184A} {479, 5184}

\bibitem[\protect\citeauthoryear{{Aubert}, {Pichon}  \& {Colombi}}{{Aubert}
  et~al.}{2004}]{Aubert2004}
{Aubert} D.,  {Pichon} C.,   {Colombi} S.,  2004, \mn@doi [\mnras]
  {10.1111/j.1365-2966.2004.07883.x}, \href
  {http://adsabs.harvard.edu/abs/2004MNRAS.352..376A} {352, 376}

\bibitem[\protect\citeauthoryear{{Baek} \& {Ferrara}}{{Baek} \&
  {Ferrara}}{2013}]{Baek2013}
{Baek} S.,  {Ferrara} A.,  2013, \mn@doi [\mnras] {10.1093/mnrasl/slt023},
  \href {https://ui.adsabs.harvard.edu/abs/2013MNRAS.432L...6B} {432, L6}

\bibitem[\protect\citeauthoryear{{Becker} \& {Bolton}}{{Becker} \&
  {Bolton}}{2013}]{Becker2013}
{Becker} G.~D.,  {Bolton} J.~S.,  2013, \mn@doi [\mnras]
  {10.1093/mnras/stt1610}, \href
  {https://ui.adsabs.harvard.edu/abs/2013MNRAS.436.1023B} {436, 1023}

\bibitem[\protect\citeauthoryear{{Behroozi}, {Wechsler}  \&
  {Conroy}}{{Behroozi} et~al.}{2013}]{Behroozi2013}
{Behroozi} P.~S.,  {Wechsler} R.~H.,   {Conroy} C.,  2013, \mn@doi [\apj]
  {10.1088/0004-637X/770/1/57}, \href
  {https://ui.adsabs.harvard.edu/abs/2013ApJ...770...57B} {770, 57}

\bibitem[\protect\citeauthoryear{{Behroozi}, {Wechsler}, {Hearin}  \&
  {Conroy}}{{Behroozi} et~al.}{2019}]{Behroozi2019}
{Behroozi} P.,  {Wechsler} R.~H.,  {Hearin} A.~P.,   {Conroy} C.,  2019,
  \mn@doi [\mnras] {10.1093/mnras/stz1182}, \href
  {https://ui.adsabs.harvard.edu/abs/2019MNRAS.488.3143B} {488, 3143}

\bibitem[\protect\citeauthoryear{{Bieri}, {Dubois}, {Rosdahl}, {Wagner}, {Silk}
   \& {Mamon}}{{Bieri} et~al.}{2017}]{Bieri2017}
{Bieri} R.,  {Dubois} Y.,  {Rosdahl} J.,  {Wagner} A.,  {Silk} J.,   {Mamon}
  G.~A.,  2017, \mn@doi [\mnras] {10.1093/mnras/stw2380}, \href
  {https://ui.adsabs.harvard.edu/abs/2017MNRAS.464.1854B} {464, 1854}

\bibitem[\protect\citeauthoryear{{Bondi}}{{Bondi}}{1952}]{Bondi1952}
{Bondi} H.,  1952, \mn@doi [\mnras] {10.1093/mnras/112.2.195}, \href
  {http://adsabs.harvard.edu/abs/1952MNRAS.112..195B} {112, 195}

\bibitem[\protect\citeauthoryear{{Boutsia}, {Grazian}, {Giallongo}, {Fiore}  \&
  {Civano}}{{Boutsia} et~al.}{2018}]{Boutsia2018}
{Boutsia} K.,  {Grazian} A.,  {Giallongo} E.,  {Fiore} F.,   {Civano} F.,
  2018, \mn@doi [\apj] {10.3847/1538-4357/aae6c7}, \href
  {https://ui.adsabs.harvard.edu/abs/2018ApJ...869...20B} {869, 20}

\bibitem[\protect\citeauthoryear{{Bouwens}, {Oesch}, {Illingworth}, {Ellis}  \&
  {Stefanon}}{{Bouwens} et~al.}{2017}]{Bouwens2017}
{Bouwens} R.~J.,  {Oesch} P.~A.,  {Illingworth} G.~D.,  {Ellis} R.~S.,
  {Stefanon} M.,  2017, \mn@doi [\apj] {10.3847/1538-4357/aa70a4}, \href
  {https://ui.adsabs.harvard.edu/abs/2017ApJ...843..129B} {843, 129}

\bibitem[\protect\citeauthoryear{{Bowler}, {McLure}, {Dunlop}, {McLeod},
  {Stanway}, {Eldridge}  \& {Jarvis}}{{Bowler} et~al.}{2017}]{Bowler2017}
{Bowler} R.~A.~A.,  {McLure} R.~J.,  {Dunlop} J.~S.,  {McLeod} D.~J.,
  {Stanway} E.~R.,  {Eldridge} J.~J.,   {Jarvis} M.~J.,  2017, \mn@doi [\mnras]
  {10.1093/mnras/stx839}, \href
  {https://ui.adsabs.harvard.edu/abs/2017MNRAS.469..448B} {469, 448}

\bibitem[\protect\citeauthoryear{{Bruzual} \& {Charlot}}{{Bruzual} \&
  {Charlot}}{2003}]{Bruzual2003}
{Bruzual} G.,  {Charlot} S.,  2003, \mn@doi [\mnras]
  {10.1046/j.1365-8711.2003.06897.x}, \href
  {https://ui.adsabs.harvard.edu/abs/2003MNRAS.344.1000B} {344, 1000}

\bibitem[\protect\citeauthoryear{{Chapon}, {Mayer}  \& {Teyssier}}{{Chapon}
  et~al.}{2013}]{Chapon2013}
{Chapon} D.,  {Mayer} L.,   {Teyssier} R.,  2013, \mn@doi [\mnras]
  {10.1093/mnras/sts568}, \href
  {http://adsabs.harvard.edu/abs/2013MNRAS.429.3114C} {429, 3114}

\bibitem[\protect\citeauthoryear{{Chardin}, {Haehnelt}, {Aubert}  \&
  {Puchwein}}{{Chardin} et~al.}{2015}]{Chardin2015}
{Chardin} J.,  {Haehnelt} M.~G.,  {Aubert} D.,   {Puchwein} E.,  2015, \mn@doi
  [\mnras] {10.1093/mnras/stv1786}, \href
  {https://ui.adsabs.harvard.edu/abs/2015MNRAS.453.2943C} {453, 2943}

\bibitem[\protect\citeauthoryear{{Chardin}, {Puchwein}  \&
  {Haehnelt}}{{Chardin} et~al.}{2017}]{Chardin2017}
{Chardin} J.,  {Puchwein} E.,   {Haehnelt} M.~G.,  2017, \mn@doi [\mnras]
  {10.1093/mnras/stw2943}, \href
  {https://ui.adsabs.harvard.edu/abs/2017MNRAS.465.3429C} {465, 3429}

\bibitem[\protect\citeauthoryear{{Circosta} et~al.,}{{Circosta}
  et~al.}{2019}]{Circosta2019}
{Circosta} C.,  et~al., 2019, \mn@doi [\aap] {10.1051/0004-6361/201834426},
  \href {https://ui.adsabs.harvard.edu/abs/2019A%26A...623A.172C} {623, A172}

\bibitem[\protect\citeauthoryear{{Costa}, {Rosdahl}, {Sijacki}  \&
  {Haehnelt}}{{Costa} et~al.}{2018}]{Costa2018}
{Costa} T.,  {Rosdahl} J.,  {Sijacki} D.,   {Haehnelt} M.~G.,  2018, \mn@doi
  [\mnras] {10.1093/mnras/sty1514}, \href
  {https://ui.adsabs.harvard.edu/abs/2018MNRAS.479.2079C} {479, 2079}

\bibitem[\protect\citeauthoryear{{Cowie}, {Barger}  \& {Trouille}}{{Cowie}
  et~al.}{2009}]{Cowie2009}
{Cowie} L.~L.,  {Barger} A.~J.,   {Trouille} L.,  2009, \mn@doi [\apj]
  {10.1088/0004-637X/692/2/1476}, \href
  {https://ui.adsabs.harvard.edu/abs/2009ApJ...692.1476C} {692, 1476}

\bibitem[\protect\citeauthoryear{{Cristiani}, {Serrano}, {Fontanot}, {Vanzella}
   \& {Monaco}}{{Cristiani} et~al.}{2016}]{Cristiani2016}
{Cristiani} S.,  {Serrano} L.~M.,  {Fontanot} F.,  {Vanzella} E.,   {Monaco}
  P.,  2016, \mn@doi [\mnras] {10.1093/mnras/stw1810}, \href
  {https://ui.adsabs.harvard.edu/abs/2016MNRAS.462.2478C} {462, 2478}

\bibitem[\protect\citeauthoryear{{Dubois}, {Devriendt}, {Slyz}  \&
  {Teyssier}}{{Dubois} et~al.}{2012}]{Dubois2012}
{Dubois} Y.,  {Devriendt} J.,  {Slyz} A.,   {Teyssier} R.,  2012, \mn@doi
  [\mnras] {10.1111/j.1365-2966.2011.20236.x}, \href
  {http://adsabs.harvard.edu/abs/2012MNRAS.420.2662D} {420, 2662}

\bibitem[\protect\citeauthoryear{{Dubois}, {Pichon}, {Devriendt}, {Silk},
  {Haehnelt}, {Kimm}  \& {Slyz}}{{Dubois} et~al.}{2013}]{Dubois2013}
{Dubois} Y.,  {Pichon} C.,  {Devriendt} J.,  {Silk} J.,  {Haehnelt} M.,  {Kimm}
  T.,   {Slyz} A.,  2013, \mn@doi [\mnras] {10.1093/mnras/sts224}, \href
  {http://adsabs.harvard.edu/abs/2013MNRAS.428.2885D} {428, 2885}

\bibitem[\protect\citeauthoryear{{Dubois}, {Volonteri}, {Silk}, {Devriendt},
  {Slyz}  \& {Teyssier}}{{Dubois} et~al.}{2015}]{Dubois2015}
{Dubois} Y.,  {Volonteri} M.,  {Silk} J.,  {Devriendt} J.,  {Slyz} A.,
  {Teyssier} R.,  2015, \mn@doi [\mnras] {10.1093/mnras/stv1416}, \href
  {https://ui.adsabs.harvard.edu/abs/2015MNRAS.452.1502D} {452, 1502}

\bibitem[\protect\citeauthoryear{{Dubroca} \& {Feugeas}}{{Dubroca} \&
  {Feugeas}}{1999}]{Dubroca1999}
{Dubroca} B.,  {Feugeas} J.,  1999, \mn@doi [Academie des Sciences Paris
  Comptes Rendus Serie Sciences Mathematiques] {10.1016/S0764-4442(00)87499-6},
  \href {http://adsabs.harvard.edu/abs/1999CRASM.329..915D} {329, 915}

\bibitem[\protect\citeauthoryear{{Federrath} \& {Klessen}}{{Federrath} \&
  {Klessen}}{2012}]{Federrath2012}
{Federrath} C.,  {Klessen} R.~S.,  2012, \mn@doi [\apj]
  {10.1088/0004-637X/761/2/156}, \href
  {http://adsabs.harvard.edu/abs/2012ApJ...761..156F} {761, 156}

\bibitem[\protect\citeauthoryear{{Ferland} et~al.,}{{Ferland}
  et~al.}{2017}]{Ferland2017}
{Ferland} G.~J.,  et~al., 2017, \rmxaa, \href
  {http://adsabs.harvard.edu/abs/2017RMxAA..53..385F} {53, 385}

\bibitem[\protect\citeauthoryear{{Finkelstein} et~al.,}{{Finkelstein}
  et~al.}{2019}]{Finkelstein2019}
{Finkelstein} S.~L.,  et~al., 2019, arXiv e-prints, \href
  {https://ui.adsabs.harvard.edu/abs/2019arXiv190202792F} {}

\bibitem[\protect\citeauthoryear{{Fragos} et~al.,}{{Fragos}
  et~al.}{2013}]{Fragos2013}
{Fragos} T.,  et~al., 2013, \mn@doi [\apj] {10.1088/0004-637X/764/1/41}, \href
  {https://ui.adsabs.harvard.edu/abs/2013ApJ...764...41F} {764, 41}

\bibitem[\protect\citeauthoryear{{Geen}, {Rosdahl}, {Blaizot}, {Devriendt}  \&
  {Slyz}}{{Geen} et~al.}{2015}]{Geen2015}
{Geen} S.,  {Rosdahl} J.,  {Blaizot} J.,  {Devriendt} J.,   {Slyz} A.,  2015,
  \mn@doi [\mnras] {10.1093/mnras/stv251}, \href
  {http://adsabs.harvard.edu/abs/2015MNRAS.448.3248G} {448, 3248}

\bibitem[\protect\citeauthoryear{{Giallongo} et~al.,}{{Giallongo}
  et~al.}{2015}]{Giallongo2015}
{Giallongo} E.,  et~al., 2015, \mn@doi [\aap] {10.1051/0004-6361/201425334},
  \href {http://adsabs.harvard.edu/abs/2015A%26A...578A..83G} {578, A83}

\bibitem[\protect\citeauthoryear{{Gnedin} \& {Abel}}{{Gnedin} \&
  {Abel}}{2001}]{Gnedin2001}
{Gnedin} N.~Y.,  {Abel} T.,  2001, \mn@doi [\na]
  {10.1016/S1384-1076(01)00068-9}, \href
  {http://adsabs.harvard.edu/abs/2001NewA....6..437G} {6, 437}

\bibitem[\protect\citeauthoryear{{Gnedin} \& {Kaurov}}{{Gnedin} \&
  {Kaurov}}{2014}]{Gnedin2014}
{Gnedin} N.~Y.,  {Kaurov} A.~A.,  2014, \mn@doi [\apj]
  {10.1088/0004-637X/793/1/30}, \href
  {https://ui.adsabs.harvard.edu/abs/2014ApJ...793...30G} {793, 30}

\bibitem[\protect\citeauthoryear{{Gnedin}, {Kravtsov}  \& {Chen}}{{Gnedin}
  et~al.}{2008}]{Gnedin2008}
{Gnedin} N.~Y.,  {Kravtsov} A.~V.,   {Chen} H.-W.,  2008, \mn@doi [\apj]
  {10.1086/524007}, \href
  {https://ui.adsabs.harvard.edu/abs/2008ApJ...672..765G} {672, 765}

\bibitem[\protect\citeauthoryear{{Grazian} et~al.,}{{Grazian}
  et~al.}{2018}]{Grazian2018}
{Grazian} A.,  et~al., 2018, \mn@doi [\aap] {10.1051/0004-6361/201732385},
  \href {https://ui.adsabs.harvard.edu/abs/2018A%26A...613A..44G} {613, A44}

\bibitem[\protect\citeauthoryear{{Guaita} et~al.,}{{Guaita}
  et~al.}{2016}]{Guaita2016}
{Guaita} L.,  et~al., 2016, \mn@doi [\aap] {10.1051/0004-6361/201527597}, \href
  {https://ui.adsabs.harvard.edu/abs/2016A%26A...587A.133G} {587, A133}

\bibitem[\protect\citeauthoryear{{Haardt} \& {Madau}}{{Haardt} \&
  {Madau}}{1996}]{Haardt1996}
{Haardt} F.,  {Madau} P.,  1996, \mn@doi [\apj] {10.1086/177035}, \href
  {http://adsabs.harvard.edu/abs/1996ApJ...461...20H} {461, 20}

\bibitem[\protect\citeauthoryear{{Habouzit}, {Volonteri}  \&
  {Dubois}}{{Habouzit} et~al.}{2017}]{Habouzit2017}
{Habouzit} M.,  {Volonteri} M.,   {Dubois} Y.,  2017, \mn@doi [\mnras]
  {10.1093/mnras/stx666}, \href
  {https://ui.adsabs.harvard.edu/abs/2017MNRAS.468.3935H} {468, 3935}

\bibitem[\protect\citeauthoryear{{Hahn} \& {Abel}}{{Hahn} \&
  {Abel}}{2011}]{Hahn2011}
{Hahn} O.,  {Abel} T.,  2011, \mn@doi [\mnras]
  {10.1111/j.1365-2966.2011.18820.x}, \href
  {http://adsabs.harvard.edu/abs/2011MNRAS.415.2101H} {415, 2101}

\bibitem[\protect\citeauthoryear{{Hartwig} et~al.,}{{Hartwig}
  et~al.}{2016}]{Hartwig2016}
{Hartwig} T.,  et~al., 2016, \mn@doi [\mnras] {10.1093/mnras/stw1775}, \href
  {https://ui.adsabs.harvard.edu/abs/2016MNRAS.462.2184H} {462, 2184}

\bibitem[\protect\citeauthoryear{{Hatfield}, {Bowler}, {Jarvis}  \&
  {Hale}}{{Hatfield} et~al.}{2018}]{Hatfield2018}
{Hatfield} P.~W.,  {Bowler} R.~A.~A.,  {Jarvis} M.~J.,   {Hale} C.~L.,  2018,
  \mn@doi [\mnras] {10.1093/mnras/sty856}, \href
  {https://ui.adsabs.harvard.edu/abs/2018MNRAS.477.3760H} {477, 3760}

\bibitem[\protect\citeauthoryear{{Hirschmann}, {Charlot}, {Feltre}, {Naab},
  {Choi}, {Ostriker}  \& {Somerville}}{{Hirschmann}
  et~al.}{2017}]{Hirschmann2017}
{Hirschmann} M.,  {Charlot} S.,  {Feltre} A.,  {Naab} T.,  {Choi} E.,
  {Ostriker} J.~P.,   {Somerville} R.~S.,  2017, \mn@doi [\mnras]
  {10.1093/mnras/stx2180}, \href
  {https://ui.adsabs.harvard.edu/abs/2017MNRAS.472.2468H} {472, 2468}

\bibitem[\protect\citeauthoryear{{Hopkins}, {Richards}  \&
  {Hernquist}}{{Hopkins} et~al.}{2007}]{Hopkins2007}
{Hopkins} P.~F.,  {Richards} G.~T.,   {Hernquist} L.,  2007, \mn@doi [\apj]
  {10.1086/509629}, \href
  {https://ui.adsabs.harvard.edu/abs/2007ApJ...654..731H} {654, 731}

\bibitem[\protect\citeauthoryear{{Howard}, {Pudritz}, {Harris}  \&
  {Klessen}}{{Howard} et~al.}{2018}]{Howard2018}
{Howard} C.~S.,  {Pudritz} R.~E.,  {Harris} W.~E.,   {Klessen} R.~S.,  2018,
  \mn@doi [\mnras] {10.1093/mnras/stx3276}, \href
  {https://ui.adsabs.harvard.edu/abs/2018MNRAS.475.3121H} {475, 3121}

\bibitem[\protect\citeauthoryear{{Hunter}}{{Hunter}}{2007}]{Hunter2007}
{Hunter} J.~D.,  2007, \mn@doi [Computing in Science and Engineering]
  {10.1109/MCSE.2007.55}, \href
  {http://adsabs.harvard.edu/abs/2007CSE.....9...90H} {9, 90}

\bibitem[\protect\citeauthoryear{{Ishigaki}, {Kawamata}, {Ouchi}, {Oguri},
  {Shimasaku}  \& {Ono}}{{Ishigaki} et~al.}{2018}]{Ishigaki2018}
{Ishigaki} M.,  {Kawamata} R.,  {Ouchi} M.,  {Oguri} M.,  {Shimasaku} K.,
  {Ono} Y.,  2018, \mn@doi [\apj] {10.3847/1538-4357/aaa544}, \href
  {https://ui.adsabs.harvard.edu/abs/2018ApJ...854...73I} {854, 73}

\bibitem[\protect\citeauthoryear{Jones, Oliphant, Peterson  et~al.}{Jones
  et~al.}{2001}]{Jones2001}
Jones E.,  Oliphant T.,  Peterson P.,   et~al., 2001, {SciPy}: Open source
  scientific tools for {Python}, \url {http://www.scipy.org/}

\bibitem[\protect\citeauthoryear{{Kakiichi} et~al.,}{{Kakiichi}
  et~al.}{2018}]{Kakiichi2018}
{Kakiichi} K.,  et~al., 2018, \mn@doi [\mnras] {10.1093/mnras/sty1318}, \href
  {https://ui.adsabs.harvard.edu/abs/2018MNRAS.479...43K} {479, 43}

\bibitem[\protect\citeauthoryear{{Kimm} \& {Cen}}{{Kimm} \&
  {Cen}}{2014}]{Kimm2014}
{Kimm} T.,  {Cen} R.,  2014, \mn@doi [\apj] {10.1088/0004-637X/788/2/121},
  \href {http://adsabs.harvard.edu/abs/2014ApJ...788..121K} {788, 121}

\bibitem[\protect\citeauthoryear{{Kimm}, {Cen}, {Devriendt}, {Dubois}  \&
  {Slyz}}{{Kimm} et~al.}{2015}]{Kimm2015}
{Kimm} T.,  {Cen} R.,  {Devriendt} J.,  {Dubois} Y.,   {Slyz} A.,  2015,
  \mn@doi [\mnras] {10.1093/mnras/stv1211}, \href
  {http://adsabs.harvard.edu/abs/2015MNRAS.451.2900K} {451, 2900}

\bibitem[\protect\citeauthoryear{{Kimm}, {Katz}, {Haehnelt}, {Rosdahl},
  {Devriendt}  \& {Slyz}}{{Kimm} et~al.}{2017}]{Kimm2017}
{Kimm} T.,  {Katz} H.,  {Haehnelt} M.,  {Rosdahl} J.,  {Devriendt} J.,   {Slyz}
  A.,  2017, \mn@doi [\mnras] {10.1093/mnras/stx052}, \href
  {http://adsabs.harvard.edu/abs/2017MNRAS.466.4826K} {466, 4826}

\bibitem[\protect\citeauthoryear{{Kulkarni}, {Worseck}  \&
  {Hennawi}}{{Kulkarni} et~al.}{2019}]{Kulkarni2019}
{Kulkarni} G.,  {Worseck} G.,   {Hennawi} J.~F.,  2019, \mn@doi [\mnras]
  {10.1093/mnras/stz1493}, \href
  {https://ui.adsabs.harvard.edu/abs/2019MNRAS.tmp.1429K} {}

\bibitem[\protect\citeauthoryear{{Laursen}, {Sommer-Larsen}  \&
  {Andersen}}{{Laursen} et~al.}{2009}]{Laursen2009}
{Laursen} P.,  {Sommer-Larsen} J.,   {Andersen} A.~C.,  2009, \mn@doi [\apj]
  {10.1088/0004-637X/704/2/1640}, \href
  {https://ui.adsabs.harvard.edu/abs/2009ApJ...704.1640L} {704, 1640}

\bibitem[\protect\citeauthoryear{{Levermore}}{{Levermore}}{1984}]{Levermore1984}
{Levermore} C.~D.,  1984, \mn@doi [\jqsrt] {10.1016/0022-4073(84)90112-2},
  \href {http://adsabs.harvard.edu/abs/1984JQSRT..31..149L} {31, 149}

\bibitem[\protect\citeauthoryear{{Livermore}, {Finkelstein}  \&
  {Lotz}}{{Livermore} et~al.}{2017}]{Livermore2017}
{Livermore} R.~C.,  {Finkelstein} S.~L.,   {Lotz} J.~M.,  2017, \mn@doi [\apj]
  {10.3847/1538-4357/835/2/113}, \href
  {https://ui.adsabs.harvard.edu/abs/2017ApJ...835..113L} {835, 113}

\bibitem[\protect\citeauthoryear{{Lusso} et~al.,}{{Lusso}
  et~al.}{2012}]{Lusso2012}
{Lusso} E.,  et~al., 2012, \mn@doi [\mnras] {10.1111/j.1365-2966.2012.21513.x},
  \href {https://ui.adsabs.harvard.edu/abs/2012MNRAS.425..623L} {425, 623}

\bibitem[\protect\citeauthoryear{{Lusso}, {Worseck}, {Hennawi}, {Prochaska},
  {Vignali}, {Stern}  \& {O'Meara}}{{Lusso} et~al.}{2015}]{Lusso2015}
{Lusso} E.,  {Worseck} G.,  {Hennawi} J.~F.,  {Prochaska} J.~X.,  {Vignali} C.,
   {Stern} J.,   {O'Meara} J.~M.,  2015, \mn@doi [\mnras]
  {10.1093/mnras/stv516}, \href
  {https://ui.adsabs.harvard.edu/abs/2015MNRAS.449.4204L} {449, 4204}

\bibitem[\protect\citeauthoryear{{Matsuoka} et~al.,}{{Matsuoka}
  et~al.}{2018}]{Matsuoka2018}
{Matsuoka} Y.,  et~al., 2018, \mn@doi [\apj] {10.3847/1538-4357/aaee7a}, \href
  {https://ui.adsabs.harvard.edu/abs/2018ApJ...869..150M} {869, 150}

\bibitem[\protect\citeauthoryear{{Matthee}, {Sobral}, {Darvish}, {Santos},
  {Mobasher}, {Paulino-Afonso}, {R{\"o}ttgering}  \& {Alegre}}{{Matthee}
  et~al.}{2017}]{Matthee2017}
{Matthee} J.,  {Sobral} D.,  {Darvish} B.,  {Santos} S.,  {Mobasher} B.,
  {Paulino-Afonso} A.,  {R{\"o}ttgering} H.,   {Alegre} L.,  2017, \mn@doi
  [\mnras] {10.1093/mnras/stx2061}, \href
  {https://ui.adsabs.harvard.edu/abs/2017MNRAS.472..772M} {472, 772}

\bibitem[\protect\citeauthoryear{{Michel-Dansac}, {Blaizot}, {Garel},
  {Verhamme}, {Kimm}  \& {Trebitsch}}{{Michel-Dansac}
  et~al.}{2020}]{MichelDansac2020}
{Michel-Dansac} L.,  {Blaizot} J.,  {Garel} T.,  {Verhamme} A.,  {Kimm} T.,
  {Trebitsch} M.,  2020, arXiv e-prints, \href
  {https://ui.adsabs.harvard.edu/abs/2020arXiv200111252M} {p. arXiv:2001.11252}

\bibitem[\protect\citeauthoryear{{Micheva}, {Iwata}  \& {Inoue}}{{Micheva}
  et~al.}{2017}]{Micheva2017}
{Micheva} G.,  {Iwata} I.,   {Inoue} A.~K.,  2017, \mn@doi [\mnras]
  {10.1093/mnras/stw1329}, \href
  {https://ui.adsabs.harvard.edu/abs/2017MNRAS.465..302M} {465, 302}

\bibitem[\protect\citeauthoryear{{Moster}, {Naab}  \& {White}}{{Moster}
  et~al.}{2018}]{Moster2018}
{Moster} B.~P.,  {Naab} T.,   {White} S. D.~M.,  2018, \mn@doi [\mnras]
  {10.1093/mnras/sty655}, \href
  {https://ui.adsabs.harvard.edu/abs/2018MNRAS.477.1822M} {477, 1822}

\bibitem[\protect\citeauthoryear{{Ocvirk} et~al.,}{{Ocvirk}
  et~al.}{2016}]{Ocvirk2016}
{Ocvirk} P.,  et~al., 2016, \mn@doi [\mnras] {10.1093/mnras/stw2036}, \href
  {https://ui.adsabs.harvard.edu/abs/2016MNRAS.463.1462O} {463, 1462}

\bibitem[\protect\citeauthoryear{{Ono} et~al.,}{{Ono} et~al.}{2018}]{Ono2018}
{Ono} Y.,  et~al., 2018, \mn@doi [\pasj] {10.1093/pasj/psx103}, \href
  {https://ui.adsabs.harvard.edu/abs/2018PASJ...70S..10O} {70, S10}

\bibitem[\protect\citeauthoryear{{Ostriker}}{{Ostriker}}{1999}]{Ostriker1999}
{Ostriker} E.~C.,  1999, \mn@doi [\apj] {10.1086/306858}, \href
  {http://adsabs.harvard.edu/abs/1999ApJ...513..252O} {513, 252}

\bibitem[\protect\citeauthoryear{{Ouchi} et~al.,}{{Ouchi}
  et~al.}{2009}]{Ouchi2009}
{Ouchi} M.,  et~al., 2009, \mn@doi [\apj] {10.1088/0004-637X/696/2/1164}, \href
  {https://ui.adsabs.harvard.edu/abs/2009ApJ...696.1164O} {696, 1164}

\bibitem[\protect\citeauthoryear{{Ouchi} et~al.,}{{Ouchi}
  et~al.}{2013}]{Ouchi2013}
{Ouchi} M.,  et~al., 2013, \mn@doi [\apj] {10.1088/0004-637X/778/2/102}, \href
  {https://ui.adsabs.harvard.edu/abs/2013ApJ...778..102O} {778, 102}

\bibitem[\protect\citeauthoryear{{Padoan} \& {Nordlund}}{{Padoan} \&
  {Nordlund}}{2011}]{Padoan2011}
{Padoan} P.,  {Nordlund} {\AA}.,  2011, \mn@doi [\apj]
  {10.1088/0004-637X/730/1/40}, \href
  {http://adsabs.harvard.edu/abs/2011ApJ...730...40P} {730, 40}

\bibitem[\protect\citeauthoryear{{Parsa}, {Dunlop}  \& {McLure}}{{Parsa}
  et~al.}{2018}]{Parsa2018}
{Parsa} S.,  {Dunlop} J.~S.,   {McLure} R.~J.,  2018, \mn@doi [\mnras]
  {10.1093/mnras/stx2887}, \href
  {https://ui.adsabs.harvard.edu/abs/2018MNRAS.474.2904P} {474, 2904}

\bibitem[\protect\citeauthoryear{{Perez} \& {Granger}}{{Perez} \&
  {Granger}}{2007}]{Perez2007}
{Perez} F.,  {Granger} B.~E.,  2007, \mn@doi [Computing in Science Engineering]
  {10.1109/MCSE.2007.53}, 9

\bibitem[\protect\citeauthoryear{{Pfister}, {Lupi}, {Capelo}, {Volonteri},
  {Bellovary}  \& {Dotti}}{{Pfister} et~al.}{2017}]{Pfister2017}
{Pfister} H.,  {Lupi} A.,  {Capelo} P.~R.,  {Volonteri} M.,  {Bellovary} J.~M.,
    {Dotti} M.,  2017, \mn@doi [\mnras] {10.1093/mnras/stx1853}, \href
  {http://adsabs.harvard.edu/abs/2017MNRAS.471.3646P} {471, 3646}

\bibitem[\protect\citeauthoryear{{Pfister}, {Volonteri}, {Dubois}, {Dotti}  \&
  {Colpi}}{{Pfister} et~al.}{2019}]{Pfister2019}
{Pfister} H.,  {Volonteri} M.,  {Dubois} Y.,  {Dotti} M.,   {Colpi} M.,  2019,
  \mn@doi [\mnras] {10.1093/mnras/stz822}, \href
  {https://ui.adsabs.harvard.edu/abs/2019MNRAS.486..101P} {486, 101}

\bibitem[\protect\citeauthoryear{{Planck Collaboration} et~al.,}{{Planck
  Collaboration} et~al.}{2016}]{Planck2015}
{Planck Collaboration} et~al., 2016, \mn@doi [\aap]
  {10.1051/0004-6361/201525830}, \href
  {http://adsabs.harvard.edu/abs/2016A%26A...594A..13P} {594, A13}

\bibitem[\protect\citeauthoryear{{Rasera} \& {Teyssier}}{{Rasera} \&
  {Teyssier}}{2006}]{Rasera2006}
{Rasera} Y.,  {Teyssier} R.,  2006, \mn@doi [\aap]
  {10.1051/0004-6361:20053116}, \href
  {http://adsabs.harvard.edu/abs/2006A%26A...445....1R} {445, 1}

\bibitem[\protect\citeauthoryear{{Ricci}, {Marchesi}, {Shankar}, {La Franca}
  \& {Civano}}{{Ricci} et~al.}{2017}]{Ricci2017}
{Ricci} F.,  {Marchesi} S.,  {Shankar} F.,  {La Franca} F.,   {Civano} F.,
  2017, \mn@doi [\mnras] {10.1093/mnras/stw2909}, \href
  {https://ui.adsabs.harvard.edu/abs/2017MNRAS.465.1915R} {465, 1915}

\bibitem[\protect\citeauthoryear{{Robertson}, {Ellis}, {Furlanetto}  \&
  {Dunlop}}{{Robertson} et~al.}{2015}]{Robertson2015}
{Robertson} B.~E.,  {Ellis} R.~S.,  {Furlanetto} S.~R.,   {Dunlop} J.~S.,
  2015, \mn@doi [\apjl] {10.1088/2041-8205/802/2/L19}, \href
  {https://ui.adsabs.harvard.edu/abs/2015ApJ...802L..19R} {802, L19}

\bibitem[\protect\citeauthoryear{{Rosdahl} \& {Teyssier}}{{Rosdahl} \&
  {Teyssier}}{2015}]{Rosdahl2015}
{Rosdahl} J.,  {Teyssier} R.,  2015, \mn@doi [\mnras] {10.1093/mnras/stv567},
  \href {http://adsabs.harvard.edu/abs/2015MNRAS.449.4380R} {449, 4380}

\bibitem[\protect\citeauthoryear{{Rosdahl}, {Blaizot}, {Aubert}, {Stranex}  \&
  {Teyssier}}{{Rosdahl} et~al.}{2013}]{Rosdahl2013}
{Rosdahl} J.,  {Blaizot} J.,  {Aubert} D.,  {Stranex} T.,   {Teyssier} R.,
  2013, \mn@doi [\mnras] {10.1093/mnras/stt1722}, \href
  {http://adsabs.harvard.edu/abs/2013MNRAS.436.2188R} {436, 2188}

\bibitem[\protect\citeauthoryear{{Rosdahl} et~al.,}{{Rosdahl}
  et~al.}{2018}]{Rosdahl2018}
{Rosdahl} J.,  et~al., 2018, \mn@doi [\mnras] {10.1093/mnras/sty1655}, \href
  {https://ui.adsabs.harvard.edu/abs/2018MNRAS.479..994R} {479, 994}

\bibitem[\protect\citeauthoryear{{Rosen} \& {Bregman}}{{Rosen} \&
  {Bregman}}{1995}]{Rosen1995}
{Rosen} A.,  {Bregman} J.~N.,  1995, \mn@doi [\apj] {10.1086/175303}, \href
  {http://adsabs.harvard.edu/abs/1995ApJ...440..634R} {440, 634}

\bibitem[\protect\citeauthoryear{{Runnoe}, {Brotherton}  \& {Shang}}{{Runnoe}
  et~al.}{2012}]{Runnoe2012}
{Runnoe} J.~C.,  {Brotherton} M.~S.,   {Shang} Z.,  2012, \mn@doi [\mnras]
  {10.1111/j.1365-2966.2012.20620.x}, \href
  {https://ui.adsabs.harvard.edu/abs/2012MNRAS.422..478R} {422, 478}

\bibitem[\protect\citeauthoryear{{Schmidt}}{{Schmidt}}{1959}]{Schmidt1959}
{Schmidt} M.,  1959, \mn@doi [\apj] {10.1086/146614}, \href
  {http://adsabs.harvard.edu/abs/1959ApJ...129..243S} {129, 243}

\bibitem[\protect\citeauthoryear{{Seiler}, {Hutter}, {Sinha}  \&
  {Croton}}{{Seiler} et~al.}{2018}]{Seiler2018}
{Seiler} J.,  {Hutter} A.,  {Sinha} M.,   {Croton} D.,  2018, \mn@doi [\mnras]
  {10.1093/mnrasl/sly122}, \href
  {https://ui.adsabs.harvard.edu/abs/2018MNRAS.480L..33S} {480, L33}

\bibitem[\protect\citeauthoryear{{Shakura} \& {Sunyaev}}{{Shakura} \&
  {Sunyaev}}{1973}]{Shakura1973}
{Shakura} N.~I.,  {Sunyaev} R.~A.,  1973, \aap, \href
  {http://adsabs.harvard.edu/abs/1973A%26A....24..337S} {24, 337}

\bibitem[\protect\citeauthoryear{{Sobral}, {Matthee}, {Darvish}, {Schaerer},
  {Mobasher}, {R{\"o}ttgering}, {Santos}  \& {Hemmati}}{{Sobral}
  et~al.}{2015}]{Sobral2015}
{Sobral} D.,  {Matthee} J.,  {Darvish} B.,  {Schaerer} D.,  {Mobasher} B.,
  {R{\"o}ttgering} H.~J.~A.,  {Santos} S.,   {Hemmati} S.,  2015, \mn@doi
  [\apj] {10.1088/0004-637X/808/2/139}, \href
  {https://ui.adsabs.harvard.edu/abs/2015ApJ...808..139S} {808, 139}

\bibitem[\protect\citeauthoryear{{Sobral} et~al.,}{{Sobral}
  et~al.}{2018}]{Sobral2018}
{Sobral} D.,  et~al., 2018, \mn@doi [\mnras] {10.1093/mnras/sty782}, \href
  {https://ui.adsabs.harvard.edu/abs/2018MNRAS.477.2817S} {477, 2817}

\bibitem[\protect\citeauthoryear{{Steidel}, {Bogosavljevi{\'c}}, {Shapley},
  {Reddy}, {Rudie}, {Pettini}, {Trainor}  \& {Strom}}{{Steidel}
  et~al.}{2018}]{Steidel2018}
{Steidel} C.~C.,  {Bogosavljevi{\'c}} M.,  {Shapley} A.~E.,  {Reddy} N.~A.,
  {Rudie} G.~C.,  {Pettini} M.,  {Trainor} R.~F.,   {Strom} A.~L.,  2018,
  \mn@doi [\apj] {10.3847/1538-4357/aaed28}, \href
  {https://ui.adsabs.harvard.edu/abs/2018ApJ...869..123S} {869, 123}

\bibitem[\protect\citeauthoryear{{Stevans} et~al.,}{{Stevans}
  et~al.}{2018}]{Stevans2018}
{Stevans} M.~L.,  et~al., 2018, \mn@doi [\apj] {10.3847/1538-4357/aacbd7},
  \href {https://ui.adsabs.harvard.edu/abs/2018ApJ...863...63S} {863, 63}

\bibitem[\protect\citeauthoryear{{Teyssier}}{{Teyssier}}{2002}]{Teyssier2002}
{Teyssier} R.,  2002, \mn@doi [\aap] {10.1051/0004-6361:20011817}, \href
  {https://ui.adsabs.harvard.edu/abs/2002A%26A...385..337T} {385, 337}

\bibitem[\protect\citeauthoryear{{Toro}, {Spruce}  \& {Speares}}{{Toro}
  et~al.}{1994}]{Toro1994}
{Toro} E.~F.,  {Spruce} M.,   {Speares} W.,  1994, \mn@doi [Shock Waves]
  {10.1007/BF01414629}, \href
  {http://adsabs.harvard.edu/abs/1994ShWav...4...25T} {4, 25}

\bibitem[\protect\citeauthoryear{{Trebitsch}, {Blaizot}, {Rosdahl}, {Devriendt}
   \& {Slyz}}{{Trebitsch} et~al.}{2017}]{Trebitsch2017}
{Trebitsch} M.,  {Blaizot} J.,  {Rosdahl} J.,  {Devriendt} J.,   {Slyz} A.,
  2017, \mn@doi [\mnras] {10.1093/mnras/stx1060}, \href
  {http://adsabs.harvard.edu/abs/2017MNRAS.470..224T} {470, 224}

\bibitem[\protect\citeauthoryear{{Trebitsch}, {Volonteri}  \&
  {Dubois}}{{Trebitsch} et~al.}{2019}]{Trebitsch2019}
{Trebitsch} M.,  {Volonteri} M.,   {Dubois} Y.,  2019, \mn@doi [\mnras]
  {10.1093/mnras/stz1280}, \href
  {https://ui.adsabs.harvard.edu/abs/2019MNRAS.tmp.1240T} {}

\bibitem[\protect\citeauthoryear{{Tremmel}, {Governato}, {Volonteri}  \&
  {Quinn}}{{Tremmel} et~al.}{2015}]{Tremmel2015}
{Tremmel} M.,  {Governato} F.,  {Volonteri} M.,   {Quinn} T.~R.,  2015, \mn@doi
  [\mnras] {10.1093/mnras/stv1060}, \href
  {https://ui.adsabs.harvard.edu/abs/2015MNRAS.451.1868T} {451, 1868}

\bibitem[\protect\citeauthoryear{{Turk}, {Smith}, {Oishi}, {Skory}, {Skillman},
  {Abel}  \& {Norman}}{{Turk} et~al.}{2011}]{Turk2011}
{Turk} M.~J.,  {Smith} B.~D.,  {Oishi} J.~S.,  {Skory} S.,  {Skillman} S.~W.,
  {Abel} T.,   {Norman} M.~L.,  2011, \mn@doi [\apjs]
  {10.1088/0067-0049/192/1/9}, \href
  {http://adsabs.harvard.edu/abs/2011ApJS..192....9T} {192, 9}

\bibitem[\protect\citeauthoryear{{Tweed}, {Devriendt}, {Blaizot}, {Colombi}  \&
  {Slyz}}{{Tweed} et~al.}{2009}]{Tweed2009}
{Tweed} D.,  {Devriendt} J.,  {Blaizot} J.,  {Colombi} S.,   {Slyz} A.,  2009,
  \mn@doi [\aap] {10.1051/0004-6361/200911787}, \href
  {http://adsabs.harvard.edu/abs/2009A%26A...506..647T} {506, 647}

\bibitem[\protect\citeauthoryear{{Volonteri}, {Reines}, {Atek}, {Stark}  \&
  {Trebitsch}}{{Volonteri} et~al.}{2017}]{Volonteri2017}
{Volonteri} M.,  {Reines} A.~E.,  {Atek} H.,  {Stark} D.~P.,   {Trebitsch} M.,
  2017, \mn@doi [\apj] {10.3847/1538-4357/aa93f1}, \href
  {http://adsabs.harvard.edu/abs/2017ApJ...849..155V} {849, 155}

\bibitem[\protect\citeauthoryear{{Wise} \& {Cen}}{{Wise} \&
  {Cen}}{2009}]{Wise2009}
{Wise} J.~H.,  {Cen} R.,  2009, \mn@doi [\apj] {10.1088/0004-637X/693/1/984},
  \href {https://ui.adsabs.harvard.edu/abs/2009ApJ...693..984W} {693, 984}

\bibitem[\protect\citeauthoryear{{Yoo}, {Kimm}  \& {Rosdahl}}{{Yoo}
  et~al.}{2020}]{Yoo2020}
{Yoo} T.,  {Kimm} T.,   {Rosdahl} J.,  2020, arXiv e-prints, \href
  {https://ui.adsabs.harvard.edu/abs/2020arXiv200105508Y} {p. arXiv:2001.05508}

\bibitem[\protect\citeauthoryear{{van Leer}}{{van Leer}}{1979}]{vanLeer1979}
{van Leer} B.,  1979, \mn@doi [Journal of Computational Physics]
  {https://doi.org/10.1016/0021-9991(79)90145-1}, 32, 101

\makeatother
\end{thebibliography}



\appendix

\section{Estimating \fesc in the simulations}
\label{sec:app:fesc}

In this appendix, we expand on footnote~\ref{fn:fesc} and explain why we choose to measure \fesc with ray-tracing rather than simply use the radiation flux propagated in the simulation.

In principle, measuring the total outward flux integrated across the virial sphere normalized by the intrinsic total luminosity of the sources within the sphere should yield an exact estimate of the ``escape fraction'', independently of the geometry of the sources within the sphere. Indeed, as (in the absence of absorption) the radiative flux decreases as $1/r^2$, the situation is essentially similar to the Gauss theorem in electrostatic: independently of the source position in a sphere, the integral of the flux will be the source luminosity. As previous studies showed that the escape of radiation is modulated on very local scales (cloud scale of a few pc, compared to the galaxy scale of a few kpc), we have $\int \bmath{F}\cdot\bmath{dS} = \fesc \dot{N}_{\rm int}$ independently of the position of the source.

\begin{figure}
  \includegraphics[width=\columnwidth]{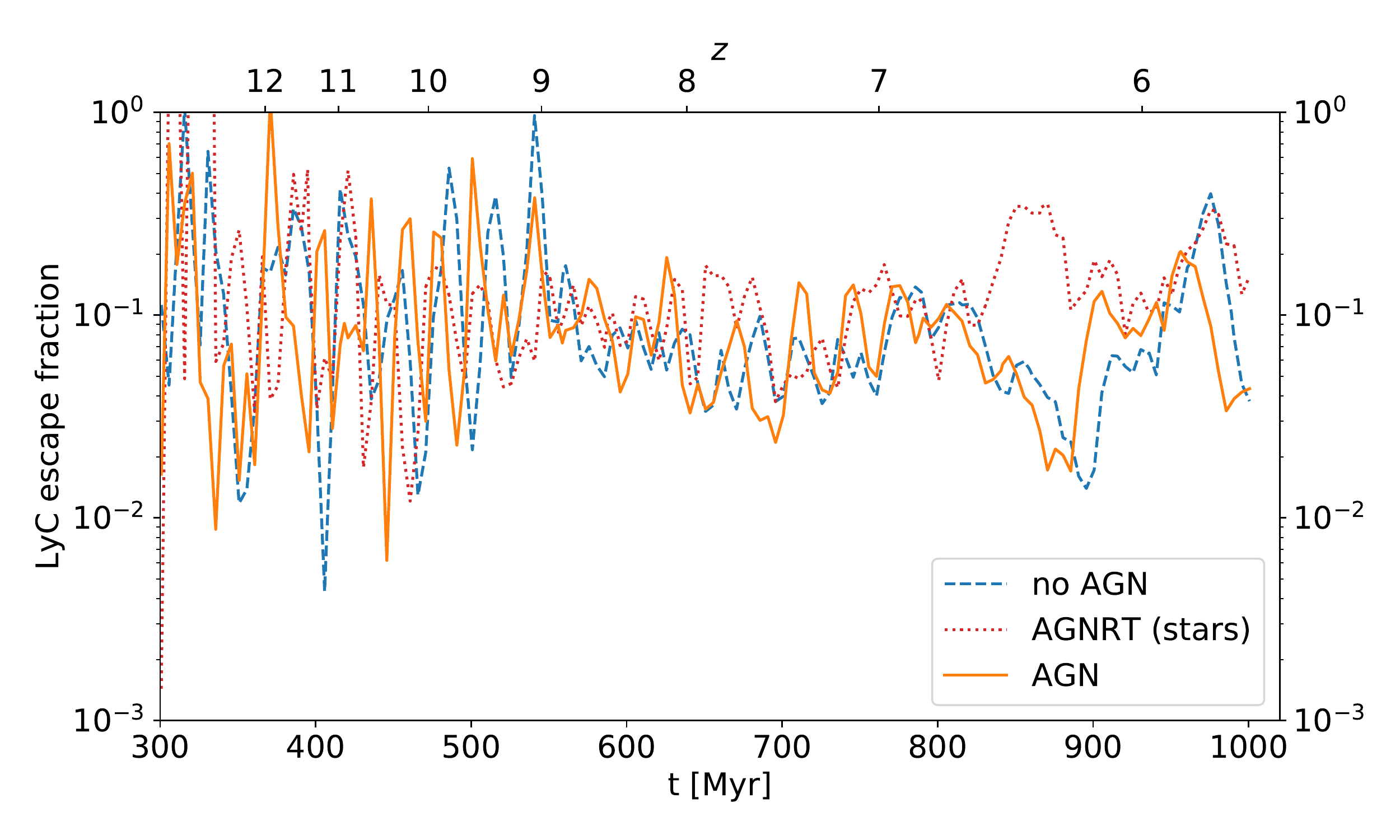}
  \caption{Escape fraction measured as the ratio of escaping to intrinsic flux for each run. Compared to Fig.~\ref{fig:fesc_vs_time}, \fesc exhibits a different behaviour in the $\sim \SI{100}{\mega\year}$ leading to the last major merger.}
  \label{fig:fesc_flux}
\end{figure}
However, this neglects the fact that moments methods are famously known to fail when radiation from two sources overlap. This is because the fluid description of radiation is unable to describe properly the crossing of two beams (they ``collide'', which is unphysical for radiation). This is in principle not a problem for the study of radiation escaping from galaxies, when we look at the radiation far away from a central, isolated sources.
Indeed, \citet{Trebitsch2017} compared the values of \fesc measured using ray-tracing and the ratio of the flux to the source luminosity and found an excellent agreement between the two methods. However, when a second bright source is found close to the central galaxy (e.g. a bright satellite), this assumption of isolation is bound to fail.
For example, Fig.~\ref{fig:fesc_flux} shows the escape fraction measured as the ratio of the outward flux to the intrinsic luminosity of the galaxies for all three runs. The three lines show a clear feature around \SIrange{800}{900}{\mega\year}, coincidental with the period between the main halo and galaxy mergers, i.e. the time during which the secondary galaxy is travelling across the central halo. The exact behaviour (increase or decrease) is only dependent on the detailed orbital configuration of the merger: we have run a test case where the AGN feedback and radiation is turned off right before the merger (at $t \sim \SI{750}{\mega\year}$), in which case the \fesc estimate follows closely the AGNRT run.

  We stress that this only affects the measure of \fesc close to the galaxy: when the distance between the two sources is very small compared to the distance at which the flux is measured, the two sources can effectively be considered as one. This means that it is only the estimate of \fesc that is affected rather than the ionizing output of the galaxy itself.
In practice, this means that the ray-traced measurement of \fesc is more robust and behaves more closely to the expectations for the ``escape fraction'' than a flux-based estimator.


\bsp	
\label{lastpage}
\end{document}